  \documentstyle[11pt]{article}
  
\newcommand{\be}{\begin{equation}}
\newcommand{\ee}{\end{equation}}
\newcommand{\bea}{\begin{eqnarray}}
\newcommand{\eea}{\end{eqnarray}}
\newcommand{\beann}{\begin{eqnarray*}}
\newcommand{\eeann}{\end{eqnarray*}}
\newcommand{\beasn}{\begin{sneqnarray}}
\newcommand{\eeasn}{\end{sneqnarray}}
\newcommand{\bref}[1]{(\ref{#1})}
\newcommand{\eps}{\epsilon}

\newcommand{\NPB}[3]{{\sl Nucl. Phys.} {\bf B#1} (19#2),  {#3}}
\newcommand{\PRD}[3]{{\sl Phys. Rev.} {\bf D#1} (19#2),   {#3}}
\newcommand{\PLB}[3]{{\sl Phys. Lett.} {\bf #1B} (19#2),  {#3}}

\newcommand{\AP}[3]{{\sl Ann. of Phys. (N.Y.)} {\bf #1} (19#2),  {#3}}
\newcommand{\IJA}[3]{{\sl Int. J. Mod. Phys.} {\bf A#1} (19#2), {#3}}
\newcommand{\CMP}[3]{{\sl Commun. Math. Phys.} {\bf #1} (19#2), {#3}}

\newcommand{\LMP}[3]{{\sl Lett. Math. Phys.} {\bf #1} (19#2), {#3}}

\catcode`@=11
\def\dif{{\rm d}}
\def\deriv{\@ifnextchar[{\@deriv}{\@deriv[]}}
   \def\@deriv[#1]#2#3{\mathchoice%
{{\dif^{#1}#2\over\dif{#3}^{#1}}}{{\dif^{#1}#2/\dif{#3}^{#1}}}%
{{\dif^{#1}#2\over\dif{#3}^{#1}}}{{\dif^{#1}#2/\dif{#3}^{#1}}}}

\def\presup#1{{}^{#1}\kern-.15em\relax}      
\def\presub#1{{}_{#1}\kern-.12em\relax}      

\def\secteqno{\@addtoreset{equation}{section}%
\def\theequation{\thesection.\arabic{equation}}}
\def\endsecteqno{\def\theequation{\@ifundefined{chapter}%
{\arabic{equation}}{\thechapter.\arabic{equation}}}}
\newcounter{subequation}
\def\thesubequation{\alph{subequation}}
\def\sneqnarray{\stepcounter{equation}\let\@currentlabel=\theequation
\setcounter{subequation}{1}
\def\@eqnnum{{\rm (\theequation\thesubequation)}}
\global\@eqcnt\z@\tabskip\@centering\let\\=\@eqncr\let\@@eqncr=\@@sneqncr
$$\halign to \displaywidth\bgroup\@eqnsel\hskip\@centering
 $\displaystyle\tabskip\z@{##}$&\global\@eqcnt\@ne
 \hskip 2\arraycolsep \hfil${##}$\hfil
 &\global\@eqcnt\tw@ \hskip 2\arraycolsep $\displaystyle\tabskip\z@{##}$\hfil
  \tabskip\@centering&\llap{##}\tabskip\z@\cr}
\def\endsneqnarray{\@@sneqncr\egroup $$\global\@ignoretrue}
\def\@@sneqncr{\let\@tempa\relax
   \ifcase\@eqcnt \def\@tempa{& & &}\or \def\@tempa{& &}
   \else \def\@tempa{&}\fi
     \@tempa \if@eqnsw\@eqnnum\stepcounter{subequation}\fi
     \global\@eqnswtrue\global\@eqcnt\z@\cr}
\def\nobiblabels{\def\@lbibitem[##1]##2{\@bibitem{##2}}}
\catcode`@=12

\def\dddot#1{\hbox{$\mathop{#1}\limits^{\ldots}$}}

\def\W{${\cal W}$}

\secteqno

\title{{\bf Particle Mechanics Models with \W-symmetries}}

\author{{\sc J. Gomis}$^\dagger$,\ %
        {\sc J. Herrero}$^\dagger$,
        {\sc K. Kamimura}$^\flat$
        {\sc and J. Roca}$^\dagger$\\
        \llap{$^\dagger$}%
        \small{\it{Departament d'Estructura i Constituents
               de la Mat\`eria}}\\
        \small{\it{Universitat de Barcelona \&}}\\
        \small{\it{Institut de F\'\i sica d'Altes Energies}}\\
        \small{\it{Diagonal, 647}}\\
        \small{\it{E-08028 BARCELONA}}\\
        \llap{$^\flat$}%
        \small{\it{Department of Physics, Toho University}}\\
        \small{\it{Funabashi}}\\
        \small{\it{274 Japan}}\\
       {\it e-mails:} \small{gomis@rita.ecm.ub.es, herrero@ecm.ub.es,}\\
                       \small{KAMIMURA@JPNYITP, roca@ecm.ub.es}}

\date{}

\begin{document}

\maketitle

\thispagestyle{empty}

\begin{abstract}
We introduce a particle mechanics model with Sp($2M$) gauge invariance.
Different partial gauge-fixings by means of sl(2) embeddings on the
gauge algebra lead to reduced models which are invariant under
diffeomorphisms and classical non-linear \W-transformations as the
residual gauge symmetries thus providing a set of
models of gauge and matter fields coupled in a \W-invariant way.
The equations of motion for the matter variables give
Lax operators in a matrix form.
We examine several examples in detail and discuss the issue of
integration of infinitesimal \W-transformations.

\end{abstract}

\vfill
\vbox{
\hfill UB-ECM-PF 93/18\null\par
\hfill TOHO-FP-9449}\null

\clearpage

\section{Introduction}
\hspace{\parindent}%

Extended conformal symmetries play an important role in two-dimensional
conformal field theories, 2d gravity models and integrable hierarchies
of non-linear differential equations. The study of non-linear extensions
of the Virasoro algebra with bosonic conformal primary fields was first
developed by Zamolodchikov \cite{Z}. Such algebras are known as \W-algebras
----- for recent reviews on \W-algebras see \cite{R,B}. Classical
\W-algebras are obtained by a contraction of \W-algebras through a
$c\to\infty$, $\hbar\to0$ limit, keeping $\hbar c$ constant.
Two methods have enjoyed much success to construct these algebras:
the Drinfel'd-Sokolov (DS) Hamiltonian reduction for Kac-Moody current
algebras \cite{DS,R1,BTD,FIG} and the zero-curvature approach
\cite{P,BFK,D,BG}.

The Kac-Moody Hamiltonian reduction consists in introducing a set of
first-class constraints in the space generated by the affine
currents $J^a(x)$ equipped with the Kac-Moody Lie-Poisson bracket:
\be
\left\{J^a(x), J^b(y) \right\}_{\rm KM} =f^{ab}_{\ \ c} \,
J^c (x) \, \delta (x - y) + \kappa \, \tilde{g}^{ab} \,
\partial_{x} \delta (x - y).
\label{km}
\ee
Here $\tilde{g}_{ab}$ is proportional to the Cartan-Killing metric and
$f_{ab}^{\ \ c}$ are the structure constants of the underlying Lie
algebra ${\cal G}$.
These constraints generate gauge transformations on the restricted space
due to its first-class nature. This gauge freedom is fixed by
introducing a second set of constraints and so a Dirac bracket can be
defined on the reduced space. The Dirac bracket algebra on this space is
the (classical) \W-algebra. The choice of the whole set of constraints
is inspired by the different inequivalent sl(2) embeddings into ${\cal
G}$.

The Kac-Moody currents generate infinitesimal transformations on the
original space via the Poisson bracket \bref{km}:
\be
\delta_a f(x) \equiv  \int \dif y
\, \epsilon_a(y) \, \left\{f(y), J^a(x) \right\}_{\rm KM}.
\label{pac}
\ee
These are the infinitesimal Kac-Moody transformations. After the
reduction the remaining currents generate infinitesimal
transformations on the reduced space via the Dirac bracket in a
similar way: the infinitesimal (classical) \W-transformations.

The zero-curvature approach starts with a ${\cal G}$-valued field
$\Lambda(x) = \Lambda^a(x)T_a$ ($T_a$ form a basis of ${\cal G}$)
transforming \`a la Yang-Mills:
\be
\delta\Lambda = \dot\beta - [\Lambda,\beta], \quad\quad
\beta(x) = \beta^a(x) T_a,
\label{zcs}
\ee
where $\beta^a (x)$ are infinitesimal parameters.
By means of a sl(2) embedding into ${\cal G}$ the components of
$\Lambda$ are partially constrained. The residual transformations
preserving these constraints are the classical \W-trans\-for\-ma\-tions.

Both approaches are equivalent since \bref{pac} becomes \bref{zcs} once
$\Lambda (t)$ is identified with the Kac-Moody holomorphic current
$J(x)$, but the second one circumvents the language of Poisson
manifolds.

In this paper we want to analyze classical \W-symmetries in the context
of particle mechanics within the zero-curvature approach. Specifically
we present a model containing gauge and matter degrees of freedom
\cite{NOS}. The transformations of the model are \bref{zcs} for the
gauge variables and ${\cal G}$ is a sp($2M$) algebra.
We call \bref{zcs} gauge transformations
because they emerge from the study of the constraint structure in the
phase space of the particle mechanics model itself, therefore
the word {\it gauge} here and throughout the rest of the paper
should be distinguished from what is usually called
gauge transformations in the context of the Kac-Moody Hamiltonian
reduction described above. According to the previous
discussion, after partially fixing this gauge freedom by means of a
sl(2) embedding we get a model which exhibits infinitesimal
transformations associated with any of the classical \W-algebras
obtainable from the $C_n$ series of Lie algebras via the DS reduction.
Furthermore, the equations of motion of the matter variables give rise
to the DS equations associated with those \W-algebras. We therefore
obtain such equations in a dynamical context. The corresponding Lax
operators are given as $M \times M$ matrices.
%

Models exhibiting \W-symmetries associated with other series of Lie
algebras can be also constructed within this framework.
For instance, if we consider some embeddings of the sl($N$) into the
sp($2N$) algebras, we obtain the transformation laws and DS
equations associated with the $A_{N-1}$ classical \W-algebras.

\medskip

The organization of the paper is as follows. In sect.\,2 we formulate
the Sp($2M$) gauge particle model for general $M$.
In sect.\,3 we consider the Sp(2) model as the simplest example and
show how reparametrization invariance arises in our model.
We also study the finite gauge transformation leading to the finite
diffeomorphism invariance of the model.
In sect.\,4 we study the Sp(4) action with its three
possible partial gauge-fixings
corresponding to the three inequivalent sl(2) embeddings in sp(4).
We integrate the gauge transformations for one of the three models
(sect.\,4.1) and perform secondary reductions of
it, ending up with systems exhibiting the non-local matrix algebra
$V_{2,2}$ \cite{bilal} and the local \W$(2,4)$ algebra which is also
associated with the principal sl(2) embedding model (sect.\,4.2).
In sect.\,5 we analyze how \W${}_3$  and \W$_3^2$
invariant models can be obtained as reductions of a sl(3) embedding
in the Sp(6) gauge particle model.
Some comments and discussions are addressed in the last section.

We also include two appendices: one about sl(2) embeddings and
DS reductions and the other about finite transformations of the
Sp$(2M)$ model before performing the gauge-fixing.


\section{Particle model with Sp$(2M)$ symmetry}
\hspace{\parindent}%

Let us consider a reparametrization-invariant model of $M$ relativistic
particles with a Sp$(2M)$ gauge group living in a Minkowskian
$d$-dimensional space-time \cite{NOS}. The dimension $d$ satisfies
$d>2M+1$ so the constraints do not trivialize the model.
The canonical action is given by
\be
\label{sp2M act}
S=\int\dif t\left(p_i\dot x_i-\lambda_{A_{ij}}\phi_{A_{ij}}\right),
\quad\quad i,j=1, \ldots ,M,\quad A=1,2,3.
\label{sp can act}
\ee
The variable $x_i^\mu(t)$ is the world-line coordinate of the
$i$-th particle and
$p_i^\mu(t)$ is its corresponding momentum. The Lagrange multipliers
$\lambda_{A_{ij}}(t)$ implement the constraints $\phi_{A_{ij}}=0$ and
satisfy

$$
\lambda_{1_{ji}} = \lambda_{1_{ij}},\quad\quad
\lambda_{3_{ji}} = \lambda_{3_{ij}}.
$$
The explicit form of $\phi_{A_{ij}}$ is
\be
\phi_{1_{ij}}=\frac 12 p_ip_j,\quad\quad
\phi_{2_{ij}}=p_ix_j\quad\quad {\rm and} \quad\quad
\phi_{3_{ij}}=\frac 12 x_ix_j.
\ee
These $2M^2+M$ constraints close under the usual Poisson bracket
$\{x_i, p_j \} = \delta_{ij}$ giving a realization of the sp($2M$)
algebra.

It is useful to introduce a matrix notation for the coordinates and
momenta of the particles
\be
R=\left(\begin{array}{c}
        r\\
        p
        \end{array}\right),
\quad\quad {\rm with} \quad\quad
r=\left(\begin{array}{c}
         x_1\\
         \vdots\\
         x_M
         \end{array}\right),\quad\quad p=\left(\begin{array}{c}
                                            p_1\\
                                            \vdots\\
                                            p_M
                                            \end{array}\right).
\ee
The conjugate of $R$ is given by
\be
\label{rbar}
\bar R=R^\top J_{2M}=   \left(p^\top,\;-r^\top\right),
\ee
where $J_{2M}$ is the $2M\times2M$ symplectic matrix
$\left(\begin{array}{cc}
              0&-1\\
             1&0
              \end{array}\right).$
The Lagrange multipliers can be written
in a form of $2M \times 2M$ symplectic
matrix

\be
\Lambda=\left(\begin{array}{cc}
              B&A\\
             -C&-B^\top
              \end{array}\right),
\ee
where the components of the $M \times M$ matrices $ A,B ,C$ are the
Lagrange multipliers $\lambda_{1_{ij}},\,\lambda_{2_{ij}},\,
 \lambda_{3_{ij}}$ respectively.

The canonical action \bref{sp can act} can be written in a matrix
form as
\be
S=\int\dif t\,\frac12\,\bar R\,{\cal D}\,R,
\ee
where ${\cal D}$ is the covariant derivative

$$
{\cal D}=\frac{\dif}{\dif t}-\Lambda.
$$
In this formulation the gauge invariance of the action is
expressed in a manifestly invariant form of Yang-Mills type%
\footnote{For a previous discussion of geometrical models with
Yang-Mills gauge theories see \cite{K}.}
with the gauge group Sp$(2M)$ \footnote{The supersymmetric version has
been studied in \cite{AG}}:
\bea
\label{sp mat tra}
&&\delta R=\beta R,
\eea
\bea
\label{sp gau tra}
&&\delta\Lambda=\dot\beta-[\Lambda,\beta],
\eea
where $\beta$ is the $2M \times 2M$ matrix form of gauge parameter
\be
\beta=\left(\begin{array}{cc}
              \beta_B & \beta_A\\
             -\beta_C & -\beta^\top_B
              \end{array}\right)
\ee
and the components $\beta_A,\beta_B,\beta_C$ are the $M \times M$
matrices associated with
the constraints $\phi_{1_{ij}},\phi_{2_{ij}},\phi_{3_{ij}}$.
The equations of motion of the matter fields are
\be
{\cal D}R=\dot R-\Lambda R=0.
\label{mat equ mot}
\ee

The infinitesimal transformation law \bref{sp gau tra} is the
compatibility condition
of the pair of equations \bref{sp mat tra} and \bref{mat equ mot}:
$$
0=[(\delta-\beta),{\cal
D}]R=-(\delta\Lambda-\dot\beta+[\Lambda,\beta])R,
$$
and it can be regarded as a zero-curvature condition.
The presence of a zero-curvature condition allows us to apply the
`soldering' \cite{P} procedure to reduce the original symmetry of the
model to
a chiral classical \W-symmetry by means of a partial gauge-fixing of the
$\Lambda$ fields. In appendix A we review this reduction method and
display the criteria for choosing the gauge-fixing.

\medskip

It is useful to express the model in terms of the Lagrangian variables.
If we write the momenta $p$ in terms of the Lagrangian variables
\be
\label{ponshell}
p=A^{-1}(\dot r-Br)\equiv K,
\ee
the action is now rewritten as
\be
S=\int\dif t\,\frac12\left(K^\top AK-r^\top Cr\right).
\label {al}
\ee
The gauge transformations become
\be
\delta r=\beta_A K+\beta_B r,\,\,\,\,\,\,\,\,\,\,
\delta\Lambda=\dot\beta-[\Lambda,\beta].
\label {open}
\ee
A characteristic feature of these Lagrangian transformations is that
the algebra is open, except for sp(2),
$$
\nonumber
[\delta_1,\delta_2]\;r=\delta_{\beta^*} r-
(\beta_A^{(2)} A^{-1} \beta_A^{(1)} -
\beta_A^{(1)} A^{-1} \beta_A^{(2)})[L]_r,
$$
\be
\\\nonumber
[\delta_1,\delta_2]\;\Lambda=\delta_{\beta^*}\Lambda,
\ee
where $\beta^*=[\beta^{(2)},\beta^{(1)}]$ and $[L]_r$ is the
Euler-Lagrange equation of motion of $r$.
There are two reasons for the appearance of an open algebra:
1) the transformations of the momenta at the
Lagrangian and Hamiltonian level do not generally coincide,
2) there are more than one first-class constraints quadratic in the
momenta.

In order to close the gauge algebra we introduce $M$ auxiliary vectors
$(F_1,\ldots,F_M)$ and modify the transformation law of the coordinates
$r$ as \be
\delta r=\beta_A (K+F)+\beta_B r.
\label{mat tra}
\ee
The transformation of $F$ is determined by the condition that $K+F$
transforms as $p$ in the Hamiltonian
 formalism. Explicitly we  get
\be
\delta F=-A^{-1}\left[\beta_A \partial_t(K+F)+
\beta_A B^\top(K+F)+(\delta A- \beta_B A)F+ \beta_A Cr\right],
\label{aux tra}
\ee
while the transformation of $\Lambda$ remains unchanged
\be
\delta\Lambda=\dot\beta-[\Lambda,\beta].
\label{mul tra}
\ee
The new algebra closes off-shell.

The invariant action under the modified gauge transformations
is
\be
\label{lagact}
S=\int\dif t \,\frac12 \left(K^\top AK-r^\top Cr-F^\top AF\right).
\ee
The redundancy of the auxiliary variables $\,F\,$ is guaranteed by the
action itself which implies $ F = 0 $ as the equation of motion.

This action is also invariant under
one-dimensional diffeomorphisms (Diff) ---$t$ reparametrizations---,
which can be obtained from the above gauge
transformations by the change of gauge parameters given in
eq.\,\bref{change para}:
\be
\beta=\tilde\beta+\epsilon\Lambda+\dot\epsilon H,
\label{new par}
\ee
where $H$ is an arbitrary element of the Cartan subalgebra of sp$(2M)$.

The Diff transformations of the fields are given by
(see appendices A and B for a derivation):
\bea
\nonumber
&\delta \Lambda^\gamma=\eps
\dot{\Lambda}^\gamma+(1+\sum_\alpha
(\alpha,\gamma)\tilde k_\alpha)\dot{\eps}\Lambda^\gamma, \quad\quad\quad
\delta \Lambda^\alpha=\eps\dot{\Lambda}^\alpha+\dot{\eps}
\Lambda^\alpha +\tilde k_\alpha\ddot{\eps},
\\\nonumber
& \delta_{\eps}r=\eps \dot{r} + \dot{\eps} Nr,
\quad\quad\quad
\delta_{\eps}F=\eps \dot{F} - \dot{\eps} NF.
\eea
These Diff transformations may be regarded as realizations
of the Virasoro group generated by (improved) Sugawara energy-momentum
tensors.
The freedom in choosing the different Virasoro realizations is reflected
in the arbitrariness of the ${\tilde k}_{\alpha}$ constants. When all of
them are zero we obtain the usual realization with all the gauge fields
having conformal weight equal to one.
The $\tilde{\beta}$ transformations are the same as in \bref{mat tra},
\bref{aux tra} and \bref{mul tra} with
$\beta_A$, $\beta_B$ and $\beta_C$ replaced by
${\tilde{\beta}}_A$, ${\tilde{\beta}}_B$ and ${\tilde{\beta}}_C$.

Once we perform a partial gauge-fixing of the $\Lambda$ matrix induced
by a sl$(2)$ embedding on sp$(2M)$, the remnant $\Lambda$ fields
will remain primaries or quasi-primaries. On the
other hand, the matter and auxiliary variables will not in general
transform as primary fields after the gauge-fixing.
This gauge-fixing procedure will be explicitly shown in the next
sections by considering several examples.
The general discussions are given in Appendices A and B.


\section{\W${}_2$ model and finite gauge transformations}
\hspace{\parindent}%

Let us now study a particle model with sl(2) gauge symmetry \cite{M}
and show how reparametrization invariance appears in our model.
Being sl(2) $\approx$ sp(2) we shall consider the model introduced in
the previous section, specialized to $M=1$.
In this case the Lagrangian gauge transformations close
off-shell and no auxiliary variables have to be introduced.
We will also construct the finite form of the residual diffeomorphism
transformations from the knowledge of the finite transformations before
the gauge-fixing. This may be a useful procedure when
direct integration of the infinitesimal residual gauge transformations
cannot be performed in a simple way.

First let us write down the model explicitly. It is described by the
first-order action \bref{sp2M act} with gauge transformations
\bref{sp mat tra} and \bref{sp gau tra}, where $R$, $\Lambda$ and
$\beta$ are given by
$$
R=\left(\begin{array}{c} x \\ p \end{array}\right),\quad\quad
\Lambda=\left(\begin{array}{rr} \lambda_2 & \lambda_1 \\
 -\lambda_3 & -\lambda_2 \end{array}\right),\quad\quad
\beta=\left(\begin{array}{rr} \beta_2 & \beta_1 \\
 -\beta_3 & -\beta_2 \end{array}\right).
$$

After the elimination of $p$ via its equation of motion the action reads
\be
S=\int\dif t\left[\frac1{2\lambda_1}(\dot
x-\lambda_2x)^2-\frac{\lambda_3}2x^2\right]
\ee
and the gauge transformations are
\bea
\nonumber
&&\delta x=\beta_2 x+\frac{\beta_1}{\lambda_1}(\dot x-\lambda_2x),
\\\nonumber
&&\delta \lambda_1=\dot\beta_1+2\lambda_1\beta_2-2\lambda_2\beta_1,
\\\nonumber
&&
\delta\lambda_2=\dot\beta_2+\lambda_1\beta_3-\lambda_3\beta_1,
\\\label{sp2 ini gau tra}
&&\delta\lambda_3=\dot\beta_3+2\lambda_2\beta_3-2\lambda_3\beta_2.
\eea
The gauge algebra still closes off-shell.
That is because, as we have mentioned in section 2, Lagrangian
open algebras can only occur in theories which possess more than one
constraint quadratic in the momenta.

\medskip

Let us now study the issues of partial gauge-fixing and remnant gauge
transformations along the lines of appendix A.
We can rewrite the matrix of Lagrange multipliers $\Lambda$ as
$$
\Lambda=\lambda_1E_++2\lambda_2h-2\lambda_3E_-,
$$
which defines the embedding of sl(2) in sp(2). In this simple case
the space of remnant fields is generated by $E_-$ alone:
${\cal G}_W={\rm Ker}\,{\rm ad}E_-=\langle E_-\rangle$, the remnant
parameter belongs to Ker$\,$ad$E_+$ and the
gauge-fixing is given by
\be
\lambda_1=1,\quad\quad\lambda_2=0,\quad\quad\lambda_3\equiv\lambda.
\label{sp2 gau fix}
\ee
The associated partially gauge-fixed action is
\be
S_{\rm pgf}=\int\dif t\left(\frac{\dot x^2}2-\lambda\frac{x^2}2\right),
\ee
which produces the matter equation of motion
\be
\ddot x+\lambda x=0.
\ee
This is precisely the DS equation $Lx=0$ where $L$ is the standard
KdV operator.

The existence of a diffeomorphism symmetry sector ---the only remnant
symmetry in this model--- can be shown by changing gauge parameters
according to \bref{new par}. In the present case ${\cal H}\cap{\cal
G}_W=\{0\}$. Hence no arbitrary constants $k_i$ can be
introduced. The redefinition \bref{new par} is here, in components,
\be
\beta_1=\eps,\quad\quad\quad\beta_2=\sigma,\quad\quad\quad
\beta_3=\rho+\lambda\eps.
\label{sp2 cha par}
\ee
If the partial gauge-fixing \bref{sp2 gau fix} is imposed then the remnant
transformations are parametrized by $\eps$ and the other parameters
are written in terms of it:
\be
\sigma=-\frac12\dot\eps,\quad\quad\quad\rho=\frac12\ddot
\eps.
\label{sp2 det par}
\ee
Using \bref{sp2 gau fix}, \bref{sp2 cha par} and \bref{sp2 det par}
in \bref{sp2 ini gau tra}
one shows that the remnant transformations are indeed (world-line)
diffeomorphisms:
\bea
\nonumber
&\delta x=\eps\dot x-\frac12\dot\eps x,
\\\label{inf res tra}
&\delta\lambda=\eps\dot\lambda+2\dot\eps\lambda+\frac12\dddot\eps.
\eea

These infinitesimal transformations can be integrated directly
to give their standard finite forms
\bea
\nonumber
&&x'(t)=(\dot f)^{-1/2}x(f(t)),
\\
&&\lambda'(t)=(\dot f)^2\lambda(f(t))+\frac 12\frac {\dot f\;\dddot f
-\frac 32 (\ddot f)^2}{(\dot f)^2}.
\label{fin res tra}
\eea

\medskip

Now we present an alternative way to find the previous finite
transformations.
We will find the finite form of the residual transformations from
the finite gauge transformations obtained before
imposing the partial gauge-fixing conditions.
First we consider a redefinition of the gauge
parameters that shows the diffeomorphism invariance before
the gauge-fixing. Next we find the finite form of these transformations.
Finally we impose the gauge conditions.
In this way we obtain the finite form of the remnant transformations
from the finite form of the transformations before the gauge-fixing.

Before imposing the gauge-fixing condition let us introduce the
following change in the gauge parameters \bref{new par}
\bea
\nonumber
\beta_1&=&\lambda_1\eps,
\\\nonumber
\beta_2&=&\sigma+\lambda_2\eps,
\\\label{sp2 new par}
\beta_3&=&\rho+\lambda_3\eps.
\eea
The gauge transformations \bref{sp2 ini gau tra} in terms of these new
parameters read:
\bea
\nonumber
\delta x&=&\eps\dot x+\sigma x,
\\\nonumber
\delta\lambda_1&=&\dot\eps\lambda_1+\eps\dot\lambda_1+2\sigma\lambda_1,
\\\nonumber
\delta\lambda_2&=&\dot\eps\lambda_2+\eps\dot\lambda_2+\dot\sigma
+\rho\lambda_1,
\\\label{sp2 new tra}
\delta\lambda_3&=&\dot\eps\lambda_3+\eps\dot\lambda_3-2\sigma\lambda_3
+\dot\rho+2\rho\lambda_2.
\eea
The $\eps$ transformation is just a
world-line reparametrization where $x$ transforms as a scalar and
the Lagrange multipliers transform as vectors.
The weights of $x$ and $\lambda$ under reparametrization
\bref{sp2 new tra} are different from the ones of after the
gauge-fixing \bref{inf res tra}. In the latter
the variable $x$ is no longer a scalar and $\lambda$ transforms as a
weight-two tensor with a ``central extension'' term.

The finite forms of the new transformations \bref{sp2 new tra} are
found for each $\epsilon,\,\sigma$ and $ \rho$ transformations:

\begin{itemize}

\item
Reparametrizations
\bea
\nonumber
x'(t)&=&x(f(t)),
\\
\lambda'_a(t)&=&\dot f(t)\;\lambda_a(f(t)),\quad\quad a=1,2,3.
\eea

\item
Local scale transformations
\bea
\nonumber
x'&=&e^\sigma x,
\\
\lambda'_1\,=\,e^{2\sigma}\lambda_1,\,\,\,\,\,\,\,\,\,\,\,\,
\lambda'_2&=&\lambda_2+\dot\sigma,\,\,\,\,\,\,\,\,\,\,\,\,
\lambda'_3\,=\,e^{-2\sigma}\lambda_3.
\eea

\item
Local redefinition of Lagrange multipliers
\bea
\nonumber
x'&=&x,
\\
\lambda'_1\,=\,\lambda_1,\,\,\,\,\,\,\,\,\,\,
\lambda'_2&=&\lambda_2+\rho\lambda_1,\,\,\,\,\,\,\,\,\,\,\,
\lambda'_3\,=\,\lambda_3+2\rho\lambda_2+\lambda_1\rho^2.
\eea

\end{itemize}

Now we impose the gauge-fixing condition \bref{sp2 gau fix}.
Notice that \bref{sp2 new par} reduces to \bref{sp2 cha par} on the
gauge-fixing surface.
Any arbitrary configuration in the gauge orbit can be realized
using a composition of the above finite transformations with
generic functions $f(t)$, $\sigma(t)$ and $\rho(t)$.
If we consider the following composition
$$
\Box\stackrel \rho \rightarrow \Box'\stackrel \sigma \rightarrow \Box''
\stackrel f \rightarrow \Box'''\equiv\stackrel{{}_{\tilde{}}}\Box,
$$
the complete finite transformation is given by
\bea
\nonumber
\tilde x&=&e^{\sigma(f(t))}x(f(t)),
\\\nonumber
\tilde\lambda_1&=&\dot f(t)e^{2\sigma(f(t))}\lambda_1(f(t)),
\\\nonumber
\tilde\lambda_2&=&\dot f(t)\left[\lambda_2(f(t))+\rho(f(t))
\lambda_1(f(t))+\dot\sigma(f(t))\right],
\\
\nonumber
\label{filam}
\tilde\lambda_3&=&\dot f(t)e^{-2\sigma(f(t))}\left[\lambda_3(f(t))
+\dot\rho(f(t))\right.
\\
&&\left.
+2\rho(f(t))\lambda_2(f(t))+\lambda_1(f(t))\rho^2(f(t))\right].
\eea
Imposing the gauge-fixing conditions \bref{sp2 gau fix} on these
transformations we obtain the finite
form of the conditions \bref{sp2 det par} for the finite gauge
parameters:
\be
\sigma(t)\,=\,-\frac12\ln\dot f(f^{-1}(t)),\,\,\,\,\,\,\,\,\,\,
\rho(t)\,=\,-\dot\sigma(t).
\ee
Using this restriction in the composition of finite gauge
transformations
\bref{filam} we arrive at the finite residual transformations
\bref{fin res tra}.
The interesting point here is that we have been able to integrate the
infinitesimal transformations \bref{inf res tra} without actually doing
it.

\section{Sp$(4)$ models}

\hspace{\parindent}%

Here we will consider the Sp$(4)$ model. In order to obtain
\W-trans\-formations we need to introduce the appropriate gauge-fixing.
For sp$(4)$ we have three different classes of sl$(2)$ embeddings
(see appendix A) which will lead to three different gauge-fixings.
Notice that not every element of these equivalence classes will
produce a gauge-fixed model written in terms of coordinates and
velocities. Only those that produce a non-singular matrix $A$ after the
gauge-fixing will have this property (see (\ref{ponshell})).

We will examine these three embeddings using the following labeling
of gauge parameters:
\be
\beta_A=\left(\begin{array}{cc}
        \beta_2&\beta_{10}\\
        \beta_{10}&\beta_5
        \end{array}\right),\quad\quad
 \beta_B=\left(\begin{array}{cc}
        \beta_3&\beta_9\\
        \beta_8&\beta_6
        \end{array}\right),\quad\quad
 \beta_C=\left(\begin{array}{cc}
        \beta_1&\beta_7\\
        \beta_7&\beta_4
        \end{array}\right).
\ee


\subsection{$(0,1)$ embedding}

\hspace{\parindent}%

Consider the gauge-fixing induced by the sl$(2)$-embedding
with characteristic $(0,1)$ given by (\ref{emb(0,1)2}).
The remnant fields after the gauge-fixing are $T$, $C$, $G$ and $H$.
Explicitly the gauge-fixing is given by
\be
\label{cccc}
\Lambda_r=\left(\begin{array}{cccc}
                H  &  0  &  0  &  1  \\
                0  & -H  &  1  &  0  \\
                C  &  \frac{T}{2}  & -H  &  0  \\
                \frac{T}{2}  &  G  &  0  &  H
              \end{array}\right).
\ee
In this gauge the action \bref{sp can act} becomes
\be
\label{swfo}
S=\int\dif t\left[(\dot x_1-H x_1)(\dot x_2+H x_2)+
\frac12\left(C x_1^2+Tx_1x_2+Gx_2^2\right)
-F_1F_2\right].
\ee
The equations of motion for the matter variables from this action are:
\be
\label{eq01}
\left(\begin{array}{c}
\;[L]_{x_1}\; \\ \;[L]_{x_2}\;
\end{array}\right)=
\left(\begin{array}{cc} C & -(\frac\dif{\dif t} + H)^2 + \frac12 T \\
                  -(\frac\dif{\dif t} - H)^2 + \frac12 T & G
\end{array}\right)
\left(\begin{array}{c} x_1 \\ x_2  \end{array}\right)\,=\,0.
\ee
These can be regarded as the DS equations for this embedding. The
corresponding Lax operator is given in a $2 \times 2$ matrix form.

There are four residual gauge transformations.
The remnant parameters live in $\ker {\rm ad} E_{+}$ and
are $\beta_{10}$, $\beta_{5}$, $\beta_{2}$ and $\beta_3 -\beta_6$.
The change given by eq.\,\bref{new par} yields the following
redefinition of these parameters after the gauge-fixing:
\bea
\nonumber
&\tilde{\beta}_2 = \beta_2, \quad\quad \tilde{\beta}_5 = \beta_5,
\quad\quad \epsilon = \beta_{10},
\\
&\alpha:=\tilde{\beta}_3 - \tilde{\beta}_6 = \beta_3 - \beta_6
-2\epsilon H - k
\dot{\epsilon}.
\eea
Here $k \equiv \frac{1}{6} (\tilde{k}_{\alpha}
-\tilde{k}_{\beta})$ is an arbitrary constant.
We keep it to stress a freedom in choosing the weight of the
fields, though it could be absorbed into the definition of $\alpha$.

The four residual transformations are:

\vskip 3mm

$\bullet \eps$-sector (Diff).
\bea
\nonumber
&\delta H=\eps\dot H+H\dot\eps+\frac{k}{2}\ddot\eps,\hskip 7mm
\delta T=\eps\dot T+2\dot\eps T-\dddot\eps,
\\\nonumber
&\delta C=\eps\dot C+(2-k)C\dot\eps, \hskip 7mm
\delta G=\eps\dot G+(2+k)G\dot\eps,
\\\nonumber
&\delta x_1=\eps\dot x_1+\frac12(k-1)x_1\dot\eps, \hskip 7mm
\delta x_2=\eps\dot x_2-\frac12(k+1)x_2\dot\eps,
\\
&\delta F_1=\eps\dot F_1-\frac12(k-1)F_1 \dot\eps, \hskip 7mm
\delta F_2=\eps\dot F_2+\frac12(k+1)F_2 \dot\eps.
\eea
In the transformations of matter and auxiliary variables we have
introduced an anti-symmetric combination of the equations of motion
(see eq.\,\bref{ineqmot diff}).

The matter and auxiliary variables $x_1$, $x_2$, $F_1$ and $F_2\,$
transform as primary fields under diffeomorphisms with weights
$\frac12(k-1)$, $-\frac12(k+1)$, $\frac12(1-k)$ and $\frac12(1+k)$
respectively.
The gauge variables $C$ and $G$ transform also as primary fields with
weights $2-k$ and $2+k$.
Instead, $T$ is a quasi-primary field with
weight $\,2\,$ and $H$ transforms as a field of weight $\,1\,$ with a
$\ddot \epsilon$ term.

$\bullet \alpha$-sector (Dilatations).
\bea
\nonumber
\delta H=\frac12 \dot\alpha,\quad\quad\delta T=0,\quad\quad
\delta C=-\alpha C,\quad\quad\delta G=\alpha G,
\\
\delta x_1=\frac12 \alpha x_1,\quad\quad\delta x_2=-\frac12 \alpha
x_2,\,\,\,\,\,\,\,\, \delta F_1=-\frac12 \alpha F_1,\quad\quad
\delta F_2=\frac12 \alpha F_2.
\eea
\vskip 4mm

$\bullet \beta_2(=\tilde{\beta}_2)$-sector.
$$
\delta H=\frac12C\beta_2,\quad\delta T=\beta_2(\dot
C-2CH)+2\dot\beta_2C, \quad\delta C=0,
$$
$$\delta G=\beta_2(4H^3-2HT-6H\dot H+\frac12\dot T+\ddot H)
-\dot\beta_2(6H^2-T-3 \dot H)+3H\ddot\beta_2-\frac12\dddot\beta_2,
$$
$$\delta x_1=\beta_2(2Hx_2+\dot x_2)-\frac12x_2\dot\beta_2+
\beta_2 F_1,\quad\delta x_2=0,
$$
\be
\delta F_1=0,\,\,\,\,\,
\delta F_2=-\beta_2\left(\dot F_1-[L]_{x_1}\right)-\frac12\dot\beta_2F_1.
\ee

$\bullet \beta_5(=\tilde{\beta}_5)$-sector.
The residual $\beta_5$ transformations can be obtained from
the $\beta_2$ transformations by the following replacements:
\be
\beta_2\leftrightarrow\beta_5,\quad\quad
H\leftrightarrow -H,          \quad\quad
C\leftrightarrow G,           \quad\quad
x_1\leftrightarrow x_2,       \quad\quad
F_1\leftrightarrow F_2.
\label{b2b5 cha}
\ee

The algebra of these residual transformations is:

$$ [\,\delta_\epsilon\,,\,\delta_{\epsilon'}\,]\,=\,\delta_{{\epsilon''}};
\hskip 7mm
{\epsilon}''\,=\,\epsilon'\,\dot\epsilon\,-\epsilon\,\dot\epsilon',
$$
$$ [\,\delta_\epsilon\,,\,\delta_{\alpha}\,]\,=\,\delta_{{\alpha}'};
\hskip 7mm
{\alpha}'=\,-\epsilon\,\dot\alpha,
$$
$$ [\,\delta_\epsilon\,,\,\delta_{\beta_2}\,]\,=\,\delta_{{\beta_2}'};
\hskip 7mm
{\beta_2}'=\,(1-k)\,\beta_2\,\dot\epsilon-\epsilon\dot\beta_2,
$$
$$ [\,\delta_\epsilon\,,\,\delta_{\beta_5}\,]\,=\,\delta_{{\beta_5}'};
\hskip 7mm
{\beta_5}'=\,(1+k)\,\beta_5\,\dot\epsilon-\epsilon\dot\beta_5,
$$
$$ [\,\delta_\alpha\,,\,\delta_{\beta_2}\,]\,=\,\delta_{{\beta_2}'};
\hskip 7mm
{\beta_2}'=\,-\alpha\,\beta_2,
$$
$$ [\,\delta_\alpha\,,\,\delta_{\beta_5}\,]\,=\,\delta_{{\beta_5}'};
\hskip 7mm
{\beta_5}'=\,\alpha\,\beta_5,
$$
$$ [\,\delta_{\beta_2}\,,\,\delta_{\beta_5}\,]\,=\,\delta_{{\epsilon}'}+
\delta_{\alpha'}+\delta_{\gamma'};
\hskip 7mm
\epsilon'\,=\,{\gamma'}\,=\,-{1\over 2}(\beta_2 \dot\beta_5-\dot\beta_2\,
\beta_5)-2\,H\,\beta_2 \beta_5,
$$
$$\alpha'=-{1\over 2}(\dot\beta_2\,\dot\beta_5-(1+k)\,\beta_2\,\ddot\beta_5
+(k-1)\beta_5\,\ddot\beta_2)+2(2+k)\beta_2\,\dot\beta_5\,H-
$$
$$
-2(2-k)\,\beta_5 \dot\beta_2\,H-2\beta_2\beta_5(\frac12 T-5H^2-k\dot H),
$$
\be
 [\,\delta_\alpha\,,\,\delta_{\alpha'}\,]\,=
   [\,\delta_\alpha\,,\,\delta_{\gamma}\,]\,=
[\,\delta_{\beta_2}\,,\,\delta_{{\beta_2}'}\,]\,=
   [\,\delta_{\beta_5}\,,\,\delta_{{\beta_5}'}\,]\,=
   [\,\delta_{\beta_2}\,,\,\delta_{\gamma}\,]\,=
   [\,\delta_{\beta_5}\,,\,\delta_{\gamma}\,]\,=\,0,
\ee
\vskip 3mm
\hspace{-\parindent}%
where the $\gamma$ transformation is a trivial transformation, i.e.
it is proportional to the equations of motion. It is explicitly given by
\bea
\nonumber
&&\delta H=\,\delta T=\,\delta C=\,\delta G=0,
\\\nonumber
&&\delta x_1=\gamma\,F_2,\,\,\,\delta x_2=\gamma\,F_1,
\\\nonumber
&&\delta F_1=-2\gamma(\dot F_1+H\,F_1)-\dot\gamma\,F_1-
\gamma[L]_{x_1},
\\\nonumber
&&\delta F_2=-2\gamma(\dot F_2-H\,F_2)-\dot\gamma\,F_2-
\gamma[L]_{x_2}.
\eea

Notice that we have an open algebra with field-dependent structure
functions.
The non-closure of the present algebra is due to the introduction
of equations of motion in the definition of the $\eps$ transformation
in terms of the original $\beta$ transformations.

\medskip

Let us present the finite forms of previous infinitesimal transformations.
The strategy is to perform the gauge-fixing on the finite
Sp$(2M)$ transformations
as it was done in the case of sl(2) (see sect.\,3).
In appendix B we display the expressions
of these finite transformations. Explicitly, we can find the
following four sets of finite transformations:

\medskip

$\bullet$ Diff. sector:\,\,
The residual finite diffeomorphisms are obtained by the following
composition of finite transformations,
($\omega:=\tilde{\beta}_3+\tilde{\beta}_6$):

$$
X\stackrel{\tilde{\beta}_7}{\longrightarrow} \Box
\stackrel{\omega}{\longrightarrow} \Box
\stackrel{{\rm diff}}{\longrightarrow}\tilde X,
$$
where $X$ stands for any variable. We can express
$\tilde{\beta}_7$ and $\omega$ in terms of
the Diff parameter $f(t)$ obtaining the residual transformations:
$$
\left\{
\begin{array}{lll}
T & \rightarrow & {\dot f}^2\, T(f) - \left( \frac{f^{(3)}}
{\dot f}- \frac{3}{2} \frac{{\ddot f}^2}{{\dot f}^2} \right)
\\
H & \rightarrow & {\dot f}\,H(f)
+ \frac{k}{2}\frac{\ddot f}{\dot f}
\\
C & \rightarrow & {\dot f}^{2-k}\,C(f)
\\
G & \rightarrow & {\dot f}^{2+k}\,G(f)
\end{array} \right.
$$

$$
\rm{matter\ variables} \left\{
\begin{array}{lll}
x_1 & \rightarrow & {\dot f}^{\frac12(k-1)}\,x_1(f)
\\
x_2 & \rightarrow & {\dot f}^{-\frac12(k+1)}\,x_2(f)
\end{array} \right.
$$

\be
\rm{auxiliary\ variables} \left\{
\begin{array}{lll}
F_1 & \rightarrow & {\dot f}^{\frac12(1-k)}\,F_1(f)
\\ \nonumber
F_2 & \rightarrow & {\dot f}^{\frac12(1+k)}\,F_2(f).
\end{array} \right.
\ee
\vskip 3mm

$\bullet \alpha$-sector (dilatations):\,\,
The finite transformations corresponding to the $\alpha$-sector,
($\,\alpha:=\tilde{\beta}_3-\tilde{\beta}_6\,$)
are the same as before and after the gauge-fixing:
$$
\left\{
\begin{array}{lll}
T & \rightarrow & T
\\
H & \rightarrow & H + \frac12{\dot \alpha}
\\
C & \rightarrow & e^{-\alpha}\, C
\\
G & \rightarrow & e^{\alpha}\, G
\end{array} \right.
$$

$$
\rm{matter\ variables} \left\{
\begin{array}{lll}
x_1 & \rightarrow & e^{\frac12\alpha}\,x_1
\\
x_2 & \rightarrow & e^{-\frac12\alpha}\,x_2
\end{array} \right.
$$

\be
\rm{auxiliary\ variables} \left\{
\begin{array}{lll}
F_1 & \rightarrow & e^{-\frac12\alpha}\,F_1
\\
F_2 & \rightarrow & e^{\frac12\alpha}\,F_2.
\end{array} \right.
\ee
\vskip 3mm

$\bullet \beta_2(=\tilde{\beta}_2)$-sector:\,\,
The residual finite $\beta_2$ transformations are obtained by the
following composition of finite transformations:
$$
X\stackrel{\tilde{\beta}_4,\tilde{\beta}_7}{\longrightarrow} \Box
\stackrel{\tilde{\beta}_9}{\longrightarrow} \Box
\stackrel{\tilde{\beta}_2}{\longrightarrow}\tilde X,
$$
and are given by:
$$
\left\{
\begin{array}{lll}
T & \rightarrow & T +
\beta_2\,\dot C + 2C\,{\dot \beta_2} -
2\beta_2\,C\,H - \frac{1}{2}{\beta_2}^2\,C^2
\\
C & \rightarrow & C
\\
H & \rightarrow & H + \frac{1}{2}\beta_2\,C
\\
G & \rightarrow & G + \beta_2\,\left(\frac12\dot T
- 6\,H\,\dot H + \ddot H - 2\,H\,T + 4\,H^3\right) +
\\
 & & + {\dot \beta_2}\,\left(T - 6\,H^2 + 3\,\dot H \right) +
3\,{\ddot \beta_2}\,H - \frac{1}{2}{\beta_2}^{(3)} +
\\
 & & + {\beta_2}^2\,\left(5\,C\,H^2
- \frac12 C\,T - 2\,C\,\dot H - 3\,\dot C \,H + \frac{1}{2}\ddot C
\right) +
\\
 & & + \beta_2\,{\dot \beta_2} \,\left(\frac{5}{2}\dot C -
8\,C\,H\right) + \frac{7}{4}{{\dot \beta_2}}^2\,C +
\frac{3}{2} \beta_2\,{\ddot \beta_2}\,C +
\\
 & & + {\beta_2}^3\left(2\,H\,C^2 - C\,\dot C \right) -
2\,{\beta_2}^2\,{\dot \beta_2} \,C^2 + \frac{1}{4}{\beta_2}^4\,C^3
\end{array} \right.
$$

$$
\rm{matter\ variables} \left\{
\begin{array}{lll}
x_1 & \rightarrow & x_1 + \beta_2 \,\left({\dot x_2}+F_1+2\,H\,x_2\right)-
\frac{1}{2}{\dot \beta_2} \,x_2 + \frac{1}{2}{\beta_2}^2\,C\,x_2
\\
x_2 & \rightarrow & x_2
\end{array} \right.
$$
\be
\rm{auxiliary\ variables} \left\{
\begin{array}{lll}
F_1 & \rightarrow & F_1
\\
F_2 & \rightarrow & F_2 -
\beta_2\,({\dot F_1}-[L]_{x_1}) -
\frac{1}{2}\dot{\beta_2}\,F_1+\frac{1}{2}{\beta_2}^2\,C\,F_1.
\end{array} \right.
\ee
\vskip 3mm

$\bullet \beta_5(=\tilde{\beta}_5)$-sector:\,\,
Again, the residual finite $\beta_5$ transformations can be
obtained from the $\beta_2$ transformations with the replacements
displayed above \bref{b2b5 cha}.
\vskip 4mm

Notice the appearance of the Schwarzian derivative in the Diff
transformation of $\,T\,$. These transformations are actually finite
symmetry transformations of the action,
under which the Lagrangian changes by a total derivative term and
the set of equations of motion remains invariant. The finite
transformations parametrized by $\alpha$, $\beta_2$ and $\beta_5$ are
a parametrization of the specific \W-transformations. One might expect,
according to the algebra of the infinitesimal transformations,
that the composition of $\beta_2$ and $\beta_5$ transformations
should give a finite Diff transformation but clearly this is not
the case.
In this sense, the above form of finite \W-transformations
is parametrized in a rather non-standard way.
\vskip 5mm

We will comment on a secondary reduction%
\footnote{A systematic study of secondary reductions of \W-algebras has
been carried out in ref.\,\cite{2red}} of this
model. When we further require $H\,=\,0$ on the gauge field
matrix \bref{cccc}
we get a system whose symmetry transformations realize a non-local
algebra discussed by Bilal \cite{bilal}.
This is expected from the form of equation of motion
\bref{eq01}.
Let us see how the residual symmetry satisfies a non-local
algebra. The condition $H=0$ further requires
\be
\delta H=\frac12 \partial (\alpha+k\dot\eps)
 +\frac12(C\beta_2+G\beta_5)\,=\,0.
\ee
If we solve it for $\alpha$, assuming a suitable boundary condition, as
\be
\label{nlalp}
(\alpha+k\dot\eps)\,=\,-\partial^{-1}(C\beta_2+G\beta_5),
\ee
there remain three residual transformations.
The fields $C$ and $G$ transform as weight 2 primaries under Diff.
They transform in a non-local way under $\beta_2$ and $\beta_5$
transformations due to \bref{nlalp}:
$$
\delta C\,=\,\partial^{-1}(\beta_2 C-\beta_5 G)C +
(\beta_5 {{\dot T}\over 2}+\dot\beta_5 T-{{\ddot \beta_5}\over 2}),
$$
$$
\delta G\,=\,-\partial^{-1}(\beta_2 C-\beta_5 G)G +
(\beta_2 {{\dot T}\over 2}+\dot\beta_2 T-{{\ddot \beta_2}\over 2}),
$$
\be
\delta T\,=\,2(\dot\beta_2 C+\dot\beta_5 G)+(\beta_2 \dot C+\beta_5 \dot G).
\ee
They are equivalent to the non-local and non-linear algebra
$V_{2,2}$ discussed in \cite{bilal}.
The matter fields are also transformed non-locally,
$$
\delta x_1\,=\,-\frac12 \partial^{-1}(C\beta_2+G\beta_5)\,x_1+
\beta_2 \dot x_2-\frac12x_2\dot\beta_2+\beta_2 F_1,
$$
$$
\delta x_2\,=\,\frac12 \partial^{-1}(C\beta_2+G\beta_5)\,x_2+
\beta_5 \dot x_1-\frac12x_1\dot\beta_5+\beta_5 F_2,
$$
$$
\delta F_1\,=\,\frac12 \partial^{-1}(C\beta_2+G\beta_5)\,F_1
-\beta_5\left(\dot F_2-[L]_{x_2}\right)-\frac12\dot\beta_5F_2,
$$
\be
\delta F_2\,=\,-\frac12 \partial^{-1}(C\beta_2+G\beta_5)\,F_2
-\beta_2\left(\dot F_1-[L]_{x_1}\right)-\frac12\dot\beta_2F_1.
\ee

\vskip 5mm

We can further impose the condition $G=1$ in addition to $H=0$.
The residual algebra becomes local again, i.e.\,$\alpha$
and $\beta_5$ are solved in local forms.
The resulting system is shown to have \W(2,4) symmetry
which will be discussed in the next subsection.


\subsection{$(1,1)$ (principal) embedding}
\hspace{\parindent}%

The gauge-fixing induced by the principal sl$(2)$ embedding
\bref{emb(1,1)} in sp$(4)$ is given by
\be
\label{lam11}
\Lambda_r=\left(\begin{array}{cccc} 0 & 0 & 0 & \sqrt{3}
       \\ 0 & 0 & \sqrt{3} & \frac15 T
       \\ \frac16 W &\frac{\sqrt{3}}{10} T & 0 & 0
       \\ \frac{\sqrt{3}}{10}T & 2 & 0 & 0
\end{array}\right).
\ee
The two remnant fields are $T$ and $W$. Here numerical factors
are taken for convention.

The action \bref{lagact} is given in this case by:
\be
\label{a11}
S_{\rm pgf}=\int\dif t\left[\frac{\dot x_1\dot x_2}{\sqrt{3}} -
                      \frac{T}{10}\left(\frac{{\dot x_1}^2}{3} -
                      \sqrt{3} x_1x_2 + F_2^2 \right) +
                      \frac{W}{12} x_1^2 + x_2^2 -
                      \sqrt{3} F_1F_2 \right].
\ee

The residual transformations are parametrized by
$\eps$ and $\rho$, related to the remnant $\beta$ parameters in the
following way:
\be
\eps - \frac{1}{20} \ddot{\rho} + \frac{3}{100} \rho T =
\frac{1}{2\sqrt{3}} \beta_{10} - \frac14 \beta_4, \quad\quad\quad
\rho = \frac16 \beta_2.
\ee
$\eps$ parametrizes the diffeomorphism sector. $T$ is a quasi-primary
weight-two field and $W$ is a primary weight-four field.  The matter
($x_i$) and auxiliary ($F_i$) fields are not primary fields.  Indeed we
can mix them to obtain a set of primary fields ($\tilde{x}_i$ and
$\tilde{F}_i$):
\bea
\nonumber
&\tilde{x}_1 = x_1, \quad\quad\quad
\tilde{x}_2 = x_2 + \frac{3}{20\sqrt{3}} T x_1 - \frac{1}{2\sqrt{3}}
\ddot{x}_1,
\\
&\tilde{F}_1 = F_1 + \frac{1}{10\sqrt{3}} T F_2, \quad\quad\quad
 \tilde{F}_2 = F_2.
\eea
The action (\ref{a11}) then becomes (neglecting total derivative terms):
\be
\label{a24}
S_{\rm pgf} = \int\dif t \left[ -\frac{1}{30} T {\dot{\tilde{x}}}_1^2 -
              \frac{1}{12} \left( {\ddot{\tilde{x}}}_1 -
                                  \frac{3}{10} T \tilde{x}_1 \right)^2 +
              \frac{1}{12} W \tilde{x}_1^2 + \tilde{x}_2^2 -
              \sqrt{3} \tilde{F}_1 \tilde{F}_2 \right].
\ee
Notice that this action is of a higher order in $\tilde{x}_1$.
Its equation of motion is
\be
\tilde x^{(4)}_1-T\ddot{\tilde x}_1-\dot T\dot{\tilde x}_1
-\left(W+\frac{3}{10}\ddot T-\frac{9}{100}T^2 \right){\tilde x_1}=0.
\ee
On the other hand, the matter
variable $\tilde{x}_2$ decouples and disappears on-shell. There are the
same number of physical degrees of freedom as in (\ref{a11}).
It has the following two residual symmetries (up to equations of motion):

\vskip 2mm

$\bullet \eps$-sector (Diff).
\bea
\nonumber
\delta T=\eps\dot T+2\dot\eps T-5\dddot\eps,\,\,\,\,\,\,\,\,\,
\delta W=\eps\dot W+4\dot\eps W,
\\\nonumber
\delta \tilde{x}_1=\eps{\dot{\tilde{x}}}_1-\frac32\tilde{x}_1\dot\eps,
\,\,\,\,\,\,\,\,
\delta \tilde{x}_2=\eps{\dot{\tilde{x}}}_2+\frac12\tilde{x}_2\dot\eps,
\\
\delta \tilde{F}_1=\eps{\dot{\tilde{F}}}_1+\frac12\tilde{F}_1\dot\eps,
\,\,\,\,\,\,\,\,\,
\delta \tilde{F}_2=\eps{\dot{\tilde{F}}}_2+\frac12\tilde{F}_2\dot\eps.
\eea

$\bullet \rho$-sector,
\bea
\nonumber
&\delta T =  3 \rho \dot{W} + 4 \dot{\rho} W,
\\\nonumber
&\delta W = -\frac{1}{20}\rho^{(7)} + \frac{7}{25}\rho^{(5)}T+
            \frac{7}{10}\rho^{(4)}\dot{T}+
\\\nonumber
&+\dddot{\rho} \left(\frac{21}{25}\ddot{T}-\frac{49}{125}T^2+
               \frac{3}{5}W \right) +
\ddot{\rho} \left( \frac{14}{25}\dddot{T}-\frac{147}{125}T\dot{T}+
               \frac{9}{10}\dot{W} \right) +
\\\nonumber
&+\dot{\rho} \left( \frac{1}{5}T^{(4)}-\frac{88}{125}T\ddot{T}-
               \frac{59}{100}\dot{T}^2+\frac{72}{625}T^3+
             \frac{1}{2}\ddot{W}-\frac{14}{25}TW  \right) +
\\\nonumber
&+\rho \left( \frac{3}{100}T^{(5)}-\frac{177}{500}\dot{T}\ddot{T}
               -\frac{39}{250}T\dddot{T}+\frac{108}{625}T^2\dot{T} +
              \frac{1}{10}\dddot{W}-
              \frac{7}{25}(TW)\dot{\ } \right),
\\\nonumber
&\delta \tilde{x}_1  = -\frac{1}{20}\dddot{\rho}{\tilde{x}}_1 +
                \frac{1}{5}\ddot{\rho} {\dot{\tilde{x}}}_1 +
                \dot{\rho} \left(
                  \frac{23}{100}T{\tilde{x}}_1 -
                  \frac{1}{2}{\ddot{\tilde{x}}}_1 \right) +
\rho \left(     -\frac{27}{100}\dot{T}{\tilde{x}}_1 -
                \frac{41}{50}T{\dot{\tilde{x}}}_1 +
                {\dddot{\tilde{x}}}_1 \right),
\\
&\delta \tilde{x}_2  =  0, \quad\quad\quad \delta \tilde{F}_1  =  0,
\quad\quad\quad \delta \tilde{F}_2  =  0.
\eea

They show that $T$ and $W$ transform according to the infinitesimal
transformations induced by the classical \W(2,4) algebra.
Thus the action (\ref{a24}) provides a particle model in which the
\W(2,4) symmetry is implemented.

\vskip 5mm

We will show how the matrix of gauge fields \bref{lam11} can
be transformed
into the form corresponding to the (0,1) embedding \bref{cccc}
with $H=0$ and $\,G=1$.
This is achieved by performing finite Sp(4) transformations
shown in Appendix B. First we make a $B_\beta$ transformation with
$$
e^{B_{\beta}}=\pmatrix{{1\over{\sqrt{6}}}&0\cr{{-1}\over{5\sqrt{6}}}T&
\sqrt{2}}
$$
in order to set the $A$ submatrix in the form
$A'=\pmatrix{0&1\cr 1&0}$. Next we realize a $C_\beta$
transformation
with
$$
{C_{\beta}}=\pmatrix{{{1}\over{5}}\dot T&0\cr 0&0}.
$$
After these gauge transformations the form of the matrix of gauge
fields becomes
\be
\label{lam24}
\Lambda''=\pmatrix{0&0&0&1\cr0&0&1&0\cr
 W  +{{4}\over{25}} T^2 -
   {{1}\over{5}}\ddot T &  \frac12 T&0&0 \cr
  \frac12 T & 1 &0&0\cr  },
\ee
This shows the equivalence of the \bref{a11} system and that given by
the action \bref{swfo} with $H=0,\,\,G=1$.
After the secondary reduction the weight 4 field $C$ is no
longer primary but is given in terms of the weight 4 primary field $W$
and the weight 2 quasi-primary field $T$ as shown in \bref{lam24}.


\subsection{$(1/2,0)$ embedding}
\hspace{\parindent}%

The gauge-fixing induced by this embedding \bref{emb(1/2,0)}  has six
remnant fields, namely $T$, $B$, $C$, $D_1$, $D_2$ and $D_3$ and is
given by
\be
\Lambda_r=\left(\begin{array}{cccc}
                0  &  0  &  1  &  0  \\
                B  & -D_1  & 0  &  D_3  \\
                \frac{T}{2}  & C  & 0  &  -B \\
                C  & D_ 2  &  0  &  D_1
              \end{array}\right).
\ee

The action \bref{sp can act} becomes:
\bea
\nonumber
S_{\rm pgf} & = & \frac12 \int\dif t \left[ {\dot x}_1^2 +
                  \frac{{\dot x}_2^2}{D_3} +
                  \frac{2}{D_3} \left( D_1 x_2 {\dot x}_2 - B x_1 {\dot x}_2 -
                  B D_1 x_1 x_2 + \,\frac12 D_1^2 x_2^2 \right)+\right.
\\
&&\left.+ \frac{B^2 x_1^2}{D_3} + \frac12 T x_1^2 + 2 C x_1 x_2
    + D_2 x_2^2 - F_1^2 - D_3 F_2^2 \right].
\eea
A characteristic feature of this embedding is that the primary field
$D_3$ appears in denominators.
The equations of motion for $x_1$ and $x_2$ are:
\be
\left(\begin{array}{c} [L]_{x_1} \cr [L]_{x_2}
\end{array}\right) \,=\,
Q \,
\left(\begin{array}{c} x_1 \cr x_2  \end{array}\right),
\ee
$$
Q \equiv \left(\begin{array}{cc}
-\frac{\dif^2}{\dif t^2} + \frac{B^2}{D_3} +
\frac12 T & -\frac{B}{D_3} \frac{\dif}{\dif t} + C - \frac{B
D_1}{D_3} \\
\frac{B}{D_3} \frac{\dif}{\dif t} + C - \frac{B D_1}{D_3} + \frac{\dot
B}{D_3} - \frac{B {\dot D}_3}{D_3^2} &
- \frac{1}{D_3} \frac{\dif^2}{\dif t^2} +
 \frac{\dot D_3}{D_3^2} \frac{\dif}{\dif t} +
D_2+\frac{D_1^2}{D_3}-\frac{\dot D_1}{D_3}+\frac{D_1 \dot
D_3}{D_3^2}
\end{array}\right).
$$

There are six residual transformations parametrized by
$\epsilon=\beta_{2}$, $\beta_{4}$, $\beta_{5}$, $\beta_{6}$,
$\beta_{9}$ and $\beta_{10}$, which are (up to equations of motion):

$\bullet \eps$-sector (Diff).
\bea
\nonumber
&\delta T=\eps\dot T+2\dot\eps T-\dddot\eps,\,\,\,\,\,\,\,\,\,\,\,\,
\delta B=\eps\dot B+(\frac 32-k)\dot\eps B,\,\,\,\,\,\,\,\,\,\,\,\,
\delta C=\eps\dot C+(\frac 32+k)\dot\eps C,
\\\nonumber
&\delta D_1=\eps\dot D_1+\dot\eps D_1+k\ddot\eps,\,\,\,\,\,\,\,
\delta D_2=\eps\dot D_2+(1+2k)\dot\eps D_2,\,\,\,\,\,\,\,
\delta D_3=\eps\dot D_3+(1-2k)\dot\eps D_3,
\\\nonumber
&\delta x_1=\eps\dot x_1-\frac12x_1\dot\eps,\quad\quad\quad
\delta x_2=\eps\dot x_2-kx_2\dot\eps,
\\
&\delta F_1=\eps\dot F_1+\frac12 F_1 \dot\eps,\quad\quad\quad
\delta F_2=\eps\dot F_2+k F_2 \dot\eps;
\eea
$k \equiv \frac{1}{6}(\tilde{k}_{\beta} -\tilde{k}_{\alpha})$
is the arbitrary constant.

\vskip 4mm
$\bullet \beta_{4}$-sector.
\bea
\nonumber
&\delta T=0,\quad\quad\delta B=0,\quad\quad\delta C=-\beta_{4} B,
\\\nonumber
&\delta D_{1}=-\beta_{4} D_3,\quad\quad\delta D_2= 2\beta_{4}
D_1-\dot\beta_{4} ,\quad\quad\delta D_3=0,
\\
&\delta x_1=0,\quad\quad\quad\delta x_2=0,\,\,\,\,\,\,\,\,\,\,\,
\delta F_1=0,\quad\quad\quad\delta F_2=0.
\eea

$\bullet \beta_{5}$-sector.
\bea
\nonumber
&\delta T=0,\quad\quad\delta B=\beta_{5}C ,\quad\quad\delta C=0,
\\\nonumber
&\delta D_{1}=-\beta_{5} D_2,\quad\quad\delta D_2= 0,\quad\quad
\delta D_3=2\beta_{5}D_1+\dot\beta_{5},
\\\nonumber
&\delta x_1=0, \quad\quad\quad\delta x_2=\beta_5 (\frac{1}{D_3} (\dot{x_2}+
D_1 x_2 - B x_1) + F_2),
\\\nonumber
&\delta F_1=0, \quad\quad\quad \delta F_2=-{\frac{1}{D_3}}\dot\beta_5 F_2-
   \frac{\beta_5}{D_3} (\dot{F_2}+D_1 F_2 - [L]_2) .
\eea

$\bullet \beta_{6}$-sector (dilatations).

\bea
\nonumber
&\delta T=0,\quad\quad\delta B=\beta_{6}B,\quad\quad\delta C=-\beta_{6} C,
\\\nonumber
&\quad\quad\delta D_{1}=-\dot\beta_{6},\quad\quad
\delta D_2= -2\beta_{6} D_2,\quad\quad\delta D_3=2\beta_{6}D_3,
\\
&\delta x_1=0,\quad\quad\quad\delta x_2=\beta_6 x_2,\,\,\,\,\,\,\,\,\,
\delta F_1=0, \quad\quad\quad \delta F_2=-\beta_6 F_2.
\eea

$\bullet \beta_{9}$-sector.

\bea
\nonumber
&\delta T=\beta_{9}(-4 B D_1+4 C D_3+ 2\dot B)+6\dot\beta_{9}B,\,\,\,\,\,\,\,\,
\delta B=-\beta_{9}\dot D_3-2\dot\beta_{9} D_3,
\\\nonumber
&\delta C= \beta_{9}(D_1^2+D_2D_3-\frac 12 T-\dot D_1)-2\dot\beta_{9} D_1
+\ddot\beta_{9},
\\\nonumber
&\delta D_1=\beta_{9} B,\quad\quad \delta D_2=-2\beta_{9} C,\quad\quad
\delta D_3=0,
\\
&\delta x_1=\beta_9 x_2,\quad\quad\delta x_2=-\beta_9 D_3 x_1,
\quad\quad
\delta F_1=\beta_9 D_3 F_2, \quad \quad\delta F_2=-\beta_9 F_1.
\eea

$\bullet \beta_{10}$-sector.

\bea
\nonumber
&\delta T=\beta_{10}(4BD_2+4CD_1+2\dot C)+6\dot\beta_{10}C,
\\\nonumber
&\delta B=-\beta_{10}(D_1^2+D_2D_3-\frac 12 T+\dot D_1)-2\dot\beta_{10}D_1
-\ddot\beta_{10},\,\,\,\,\,\,\,\,\,\,
\delta C= \beta_{10}\dot D_2+2\dot\beta_{10} D_2,
\\\nonumber
&\delta D_1=-\beta_{10} C,\quad\quad \delta D_2=0,\quad\quad
\delta D_3=-2\beta_{10} B,
\\\nonumber
&\delta x_1=\beta_{10} (\frac{1}{D_3} (\dot{x_2} + D_1 x_2 - B x_1) + F_2),
\,\,\,\,\,\,\,\,\,\,
\delta x_2=\beta_{10} (\dot{x_1} - D_1 x_1 + F_1) - \dot{\beta_{10}} x_1,
\\\nonumber
&\delta F_1=-\beta_{10}(\dot F_2-D_1 F_2-[L]_2),\,\,\,\,\,\,\,\,\,
\\
&\delta F_2=- \dot\beta_{10}\frac{F_1}{D_3}-\frac{\beta_{10}}{D_3}
  (\dot F_1+D_1 F_1-B F_2-[L]_1).
\eea

The three $D_i$ fields are the generators of the $\beta_6$,
$\beta_5$
and $\beta_4$ gauge transformations. These transformations close between
themselves for the $D_i$ fields forming a sl(2) Kac-Moody algebra.


\section{Sl$(3)$ models}
\hspace{\parindent}%

The \W$(2,4,\ldots,2M)$ gauge transformations can be obtained by
considering the principal sl(2) embedding in a general sp$(2M)$ algebra.
It is also possible to construct particle-like models having
symmetries related to other ${\cal W}$ algebras. If we want to obtain
models related to the $A_{N-1}$ series (for instance, the
${\cal W}_N$ algebras) we have to look for embeddings of
the sl$(N)$ algebras in the symplectic algebras. There is a canonical
embedding, namely,
\be
{\rm sl}(N) \oplus {\rm u}(1) \subset {\rm sp}(2N).
\ee
Explicitly, the set of matrices of the form
\be
\left(\begin{array}{cc}
                B & 0 \\ 0 & -B^\top \end{array}\right)
\ee
are a subalgebra of sp$(2N)$ isomorphic to sl$(N) \oplus$  u(1).
We may
construct the particle model taking this specific form of the gauge fields
matrix but we cannot follow the procedure outlined in section 3
because $A=0$ so eq.\,(\ref{ponshell}) is no longer valid
to eliminate the $p$ variables through their equations of motion.
Indeed, when we put all the momenta on-shell,
we obtain a null Lagrangian.

However we can deal with other embeddings and obtain particle-like actions.
For example, let us consider a canonical action (\ref{sp can act}) with
$M=3$  taking $\Lambda$ as:
\be
\Lambda = \left(\begin{array}{cccccc}
              \lambda_7 & 0 & 0 & 0 & \lambda_1 & \lambda_3 \\
              0 & \lambda_7-\lambda_8 & \lambda_5 & \lambda_1 & 0 & 0 \\
              0 & \lambda_2 & \lambda_8 & \lambda_3 & 0 & 0 \\
              0 & \lambda_4 & \lambda_6 & -\lambda_7 & 0 & 0 \\
              \lambda_4 & 0 & 0 & 0 & \lambda_8-\lambda_7 & -\lambda_2 \\
              \lambda_6 & 0 & 0 & 0 & -\lambda_5 & -\lambda_8
          \end{array}\right).
\ee
We are considering only a part of the $\phi_{A_{ij}}$ constraints
(see appendix A).
They still close under Poisson bracket giving a realization of the
sl$(3)$ algebra
(we then have a sl$(3)$ subalgebra of sp$(6)$). The gauge
transformations
still are (\ref{sp gau tra}) and (\ref{sp mat tra}) with the following
$\beta$ matrix:
\be
\beta = \left(\begin{array}{cccccc}
     \beta_7 & 0 & 0 & 0 & \beta_1 & \beta_3 \\
     0 & \beta_7-\beta_8 & \beta_5 & \beta_1 & 0 & 0 \\
     0 & \beta_2 & \beta_8 & \beta_3 & 0 & 0 \\
     0 & \beta_4 & \beta_6 & -\beta_7 & 0 & 0 \\
     \beta_4 & 0 & 0 & 0 & -\beta_7+\beta_8 & -\beta_2 \\
     \tilde{\beta_6} & 0 & 0 & 0 & -\beta_5 & -\beta_8
\end{array}\right) = \tilde{\beta} + \eps \Lambda.
\ee


Let us perform the following gauge-fixing (induced by the principal
sl$(2)$ embedding in sl$(3)$):
\bea
\nonumber
& \lambda_1 = \lambda_2 = 1, \quad\quad\quad
\lambda_3 = \lambda_7 = \lambda_8 = 0,
\\
& \lambda_4 = \lambda_5 = \frac12 T, \quad\quad\quad
\lambda_6 =  W.
\eea
Again we cannot use (\ref{ponshell}) because $\det A = 0$ but we can
get a (higher order) particle-like Lagrangian once we eliminate the
momenta variables:
\be
S_{\rm pgf} = \int\dif t \left[ \frac12 \left(\ddot x_1 \dot x_3
-\dot x_1 \ddot x_3 \right) - \frac12 T \left( x_1 \dot x_3 -
              \dot x_1 x_3 \right) -W x_1 x_3 \right].
\ee
The equations of motion for the $x_i$ variables are:
\bea
\nonumber
[L]_{x_1} = \dddot x_3 - T \dot x_3 - (W+\frac12 \dot{T})x_3,
\\
\label{sl3 x mot}
[L]_{x_3} =-\dddot x_1 + T \dot x_1 + (-W+\frac12 \dot{T})x_1.
\eea
They are two copies of the DS equation for \W$_3$.

This action exhibits a (classical) ${\cal W}_3$ symmetry, being the
remnant parameters $\eps$ and $\rho = \tilde{\beta}_3$:

$\bullet \eps$-sector (Diff).
\bea
\nonumber
& \delta T = \eps \dot{T} + 2 \dot{\eps} T -2 \dddot{\eps},
\\\nonumber
& \delta W = \eps \dot{W} + 3 \dot{\eps} W,
\\\nonumber
& \delta x_1 = \eps \dot x_1 - \dot{\eps} x_1,
\\
\label{sl3 x tra1}
& \delta x_3 = \eps \dot x_3 - \dot{\eps} x_3.
\eea

$\bullet \rho$-sector.
\bea
\nonumber
& \delta T = 2 \rho \dot{W} + 3 \dot{\rho} W,
\\\nonumber
& \delta W = \frac16 \rho^{(5)} - \frac56 \dddot{\rho} T -
            \frac{5}{4} \ddot{\rho} \dot{T} +
   \dot{\rho} \left(- \frac{3}{4} \ddot{T} + \frac{2}{3} T^2 \right) +
    \rho \left( -\frac{1}{6} \dddot{T} + \frac{2}{3} T \dot{T} \right),
\\\nonumber
& \delta x_1 = -\frac{1}{2} \left( \frac{1}{3} \ddot{\rho} x_1
- \dot{\rho} \dot x_1 + 2 \rho \ddot x_1 - \frac{4}{3} \rho T x_1
\right),
\\
\label{sl3 x tra2}
& \delta x_3 = \frac{1}{2} \left( \frac{1}{3} \ddot{\rho} x_3
- \dot{\rho} \dot x_3 + 2 \rho \ddot x_3 - \frac{4}{3} \rho T x_3
\right).
\eea

\medskip
We can also present a model exhibiting the symmetry associated with the
\W-algebra generated through the only non-principal sl$(2)$ embedding
into sl$(3)$, namely \W$_3^2$. By performing the following
gauge-fixing:
\bea
\nonumber
& \lambda_1 = \lambda_2 = 0, \quad\quad\quad
\lambda_3 = -1,
\\
& \lambda_4 = B, \quad\quad\quad
\lambda_5 = C, \quad\quad\quad
\lambda_6 = -T, \quad\quad\quad
\lambda_7 = -\lambda_8 = H,
\eea
we obtain the following action:
\bea
\nonumber
& S_{\rm pgf} = \int\dif t \frac{1}{C}\left[
H x_1 \ddot x_2 - \dot x_1 \ddot x_2 + \left(H + \frac{\dot C}{C}\right)
\dot x_1 \dot x_2 + 2 \left(\dot H - \frac{H \dot C}{C} + H^2 \right)
\dot x_1 x_2 \right.
\\
&\quad \left. - \left(T + \frac{H \dot C}{C} + H^2 \right) x_1 \dot x_2 +
2H \left(T - \dot H + \frac{H \dot C}{C} - H^2 \right) x_1 x_2 +
B C x_1 x_2 - \right].
\eea
The equations of motion for $x_1$ and $x_2$ provide DS equations for
\W$_3^2$:
\bea
\nonumber
& \dddot x_2 - \frac{2 \dot C}{C} \ddot x_2 + \left( \frac{2H\dot C}{C}
- 3\dot H - 3 H^2 - T - \frac{\ddot C}{C} + \frac{2 {\dot C}^2}{C^2}
\right) \dot x_2 +
\\
& + \left( \frac{4\dot H \dot C}{C} + \frac{4 H^2 \dot C}{C} + \frac{2 H
\ddot C}{C} - \frac{4 H {\dot C}^2}{C^2} - 6 H \dot H - 2 H^3 + 2HT + BC
- 2 \ddot H \right) x_2 = 0,
\\\nonumber
& \dddot x_1 - \frac{\dot C}{C} \ddot x_1 + \left( \frac{2H\dot C}{C}
- 3\dot H - 3 H^2 - T \right) \dot x_1 +
\\
& + \left( \frac{\dot H \dot C}{C} - \frac{H^2 \dot C}{C}
+ 2 H^3 - 2HT - BC - \ddot H - \dot T + \frac{T \dot C}{C} \right) x_1 =
0.
\eea

The previous action exhibits \W$_3^2$ symmetry:

$\bullet \eps$-sector (Diff).
\bea
\nonumber
&\delta H=\eps\dot H+H\dot\eps+ k \ddot\eps,
\\\nonumber
&\delta T=\eps\dot T+2\dot\eps T-\frac12\dddot\eps,
\\\nonumber
&\delta B=\eps\dot B+(\frac32-3k)B\dot\eps,
\\\nonumber
&\delta C=\eps\dot C+(\frac32+3k)C\dot\eps,
\\\nonumber
&\delta x_1=\eps\dot x_1+(k-\frac12)x_1\dot\eps,
\\\nonumber
&\delta x_2=\eps\dot x_2+2k x_2\dot\eps.
\eea

$\bullet \alpha(=-\frac12\beta_5)$-sector.
\bea
\nonumber
& \delta H=\dot\alpha,\quad\quad\delta T=0,\quad\quad
\delta B=-3\alpha B,\quad\quad\delta C=3\alpha C,
\\
& \delta x_1= \alpha x_1,\quad\quad\delta x_2=2 \alpha x_2.
\eea

$\bullet \beta_2(=\tilde{\beta}_2)$-sector.
\bea
\nonumber
& \delta H=\frac12B\beta_2,\quad\delta T=\beta_2(\frac12 \dot
B-3BH)+\frac32\dot\beta_2B, \quad\delta B=0,
\\\nonumber
& \delta C=\beta_2(9H^2-3\dot H - T) - 6 \dot \beta_2 H + \ddot \beta_2,
\\\nonumber
& \delta x_1=\beta_2\left( \frac{1}{C} (H^2-T-\dot H) x_1 - \frac{2H}{C}
\dot x_1 + \ddot x_1 \right),
\\
& \delta x_2 = \frac{\beta_2}{C} \left( (8H^2 - \frac{2H\dot C}{C} + 2
\dot H) x_2 + (\frac{\dot C}{C} - 2H)\dot x_2 - \ddot x_2 \right) +
\frac{\dot \beta_2}{C} (\dot x_2 - 2H x_2).
\eea

$\bullet \beta_6(=\tilde{\beta}_6)$-sector.
\bea
\nonumber
& \delta H=-\frac12C\beta_6,\quad\delta T=\beta_6(\frac12 \dot
C+3CH)+\frac32\dot\beta_6C, \quad\delta C=0,
\\\nonumber
& \delta B=\beta_6(T -9H^2-3\dot H) - 6 \dot \beta_6 H - \ddot
\beta_6,
\\
& \delta x_1=0, \quad\quad\quad \delta x_2=0.
\eea

\section{Conclusions}
\hspace{\parindent}%

We have introduced a particle mechanics model in phase space which
can be recast as a one-dimensional Sp($2M$) gauge theory.
Different partial gauge-fixings of this model by sl$(2)$ embeddings
in sp$(2M)$ yield reduced theories in which the remnant Lagrange
multiplier variables correspond to generators of classical
\W-algebras associated with the $C_n$ simple Lie algebras.
We have also shown how to obtain models invariant under
\W-algebras related to other series, such as the $A_n$.

In relation with the issue of finite \W-transformations,
the simplest reduced theory, with ${\cal W}_2$ symmetry, has only
one remnant Lagrange multiplier which transforms as a weight-two
quasi-primary field.
In this case, the finite form of its symmetry transformations is
easily obtained by using the finite transformations of the sp(2)
model and restricting them to those satisfying the gauge-fixing
condition.

Application of this procedure to the model associated
with the $(0,1)$ sl$(2)$ embedding in sp$(4)$ yields
finite symmetry transformations of its action.
These finite transformations are perfectly acceptable as a
parametrization of the gauge freedom of the system and
they are actually useful for building the general solution
of the model.
However they cannot be regarded as standard finite \W-diffeomorphism
transformations because their composition does not give
ordinary diffeomorphisms. In order to obtain the expected form of finite
\W-diffeomorphism transformations
one might introduce a non-linear change of infinitesimal gauge parameters
before the gauge-fixing by modifying the Yang-Mills transformations in a
similar way as it was done to extract the ordinary diffeomorphism.
Indeed, other approaches \cite{wgeo5,NOS2} seem to point in the
direction of treating all \W-transformations as Diff of an extended
space.

We have also shown a derivation of a non-local algebra in the course of
a secondary reduction of the Sp$(4)$ model.
This secondary reduction does not come from a sl(2) embedding in sp(4)
because the non-remnant gauge parameter is not solved algebraically.
However the non-local algebra \cite{bilal} appears in the dynamical
context.
It may be interesting to study various non-local algebras
arising from the secondary reductions of Sp(2M) models.

\vskip 5mm

{\bf Acknowledgements}

\hspace{\parindent}%
We thank J.M.\,Figueroa-O'Farrill and E.\,Ramos for a discussion about
\W-Diff.
J.\,H.\,acknowledges a fellowship from Generalitat de Catalunya and
J.\,G.\,is grateful to Ministerio de Educaci\' on y Ciencia of Spain
for a grant.
This work is partially supported by Robert A.\,Welch Foundation,
NSF Grant PHY 9009850, NATO Collaborative
Research Grant (0763/87) and CICYT project no.\ AEN93-0695.

\vskip 10mm

\appendix

\section{Zero-curvature condition and  $\cal W$-transformations}
\hspace{\parindent}%

In this appendix we review the `soldering' procedure to construct
classical chiral \W-transformations \cite{P} and its relation with
the zero-curvature approach and sl($2,R$)-embedding technique
\cite{BFK,BG,D}.

Let $\Lambda (t)$ be a Lie-algebra valued field transforming
\be
\delta\Lambda=\dot\beta-[\Lambda,\beta].
\ee
This can be regarded as a zero-curvature condition:
\be
\left[ \delta - \beta, \partial_t - \Lambda \right] = 0
\ee
Let us now consider a partial gauge-fixing on the matrix of
Lagrange multipliers $\Lambda (t)$:
\be
\label{pgf}
\Lambda=M+W,
\ee
where $M$ is a non-zero constant element of ${\cal G}$ and $ W=W^b T'_b$.
The Lie algebra elements $T'_b$ span ${\cal G}_W$, a subspace of
${\cal G}$ $(b=1,\ldots,\dim{\cal G}_W < \dim{\cal G})$ and
$W^b$ are the remnant fields living in ${\cal G}_W$.
We are looking for residual gauge transformations \bref{sp gau tra}
preserving the partial gauge-fixing \bref{pgf}.
The zero-curvature condition is now the gauge-slice conservation
condition:
\be
\label{zeroc}
\left[ M,\beta \right] + \delta W = \dot{\beta} - \left[ W, \beta
\right].
\ee
The possible partial gauge-fixings, i.e. the choices of $M$ and $W$,
are restricted once we impose the following two requirements:

\begin{itemize}
\item
We want to express a subset of the gauge parameters
$\beta^a$ as a function of another subset (remnant parameters) and
the remnant fields $W^b$ in a purely algebraic way.

\item
The residual transformations should include a diffeomorphism (Virasoro)
sector in such a way that we could identify a weight-two quasi-primary
field to it.
\end{itemize}

The first requirement is algebraically equivalent to the condition of
the total set of constraints being second-class in the Kac-Moody
Hamiltonian reduction.
Both requirements are satisfied if the partial gauge-fixing \bref{pgf}
is induced by a sl$(2,R)$ embedding \cite{BTD,R2},
$\cal S$, of the original Lie algebra ${\cal G}$,
\be
\label{lgf}
M=E_+, \quad \quad \quad \quad {\cal G}_W = \ker {\rm ad}E_-,
\ee
where $E_+$, $E_-$ and $h$ are the defining elements of the sl$(2,R)$
embedding:
\be
\left[ h, E_{\pm} \right] = \pm E_{\pm}, \quad \quad \quad \left[ E_+,
E_- \right] = h.
\ee
The mapping ad$\cal S$ given by ${\rm ad} {\cal S}:\;a\rightarrow
{\rm ad}a $ where $a\in{\cal S}$ and
\bea
\nonumber
&& {\rm ad}a:  {\cal G}\longrightarrow {\cal G}
\\\nonumber
&&\quad\quad g\longrightarrow [a,g],  \quad\quad a\in{\cal G},
\eea
is a representation of $\cal S$ on $\cal G$.
This representation is completely reducible so $\cal G$ (as a vector
space) decomposes to a direct sum of invariant subspaces of spin $j$
(integer or half-integer) and multiplicity $n_j$ ({\it branching}):
\be
{\cal G} = \sum_{j \geq 0} \sum_{i = 1}^{n_j} \dot{+} \ {\cal
G}_j^{(i)}, \quad \quad
{\cal G}_j^{(i)} = \sum_{m = -j}^{j} \dot{+} \ {\cal G}_{j,m}^{(i)},
\quad \quad
\sum_{j \geq 0} n_j (2j+1) = \dim \cal G.
\ee
The ${\cal G}_{j,m}^{(i)}$ are one-dimensional eigenspaces of ad$h$
with eigenvalue $m$.
A spin $1$ subspace is always present in the branching, namely, ${\cal S}$
itself (denoted by ${\cal G}_{\underline 1}$).
They define a gradation of $\cal G$:
\bea
\nonumber
&\hat{\cal G}_m = \left\{ \begin{array}{ll}
        {\displaystyle \sum_{j \geq m} \sum_{i=1}^{n_j} \dot{+} \ {\cal
G}_{j,m}^{(i)}},
      & \mbox{if $m$ is an eigenvalue of ad$h$} \\ [5mm]
        \left\{ 0 \right\}, & \mbox{otherwise}
      \end{array} \right.
\\\nonumber
\\
&\Rightarrow \quad \quad {\cal G} = \sum_m \dot{+} \ \hat{\cal G}_m,
\quad \quad
\left[ \hat{\cal G}_m, \hat{\cal G}_n \right] \subset \hat{\cal G}_{m+n}.
\eea

According to \bref{lgf}, every remnant field lives in
$\hat{\cal G}_{m=-j}$ and there are
\be
N({\cal S}) = \sum_{j \geq 0} n_j
\ee
such fields.

The presence of this gradation ensures that the first requirement is
satisfied. Indeed, remnant parameters live in $\ker{\rm ad}E_+$, i.e.
in $\hat{\cal G}_{m=j}$. Restrictions of the zero-curvature condition
(\ref{zeroc}) to subspaces $\hat{\cal G}_j$ allow us to express in an
algebraic way parameters living in $\hat{\cal G}_{j-1}$ as functions of
parameters living in $\hat{\cal G}_j$ (and fields) because $M$ lives in
$\hat{\cal G}_1$. So, as we go down on the spectrum of $m$'s, we have an
algebraic algorithm to express all the gauge parameters as functions of
those living in $\ker{\rm ad}E_+$ and fields. Finally, restrictions of
(\ref{zeroc}) to subspaces $\hat{\cal G}_{m=-j}$ give the transformations
of the remnant fields, $\delta W^b$.

\medskip

The existence of a Virasoro sector can be shown by performing
a decomposition of parameters:
$\beta \rightarrow \tilde{\beta},\epsilon$.
Consider the following change:
\be
\beta = \tilde{\beta} + \eps \Lambda + \dot{\eps} H,
\label{change para}
\ee
where $H=\sum_\alpha\tilde k_\alpha H_\alpha$ is a general element of
the Cartan subalgebra ${\cal H}$ of ${\cal G}$, with constant
coefficients and $\tilde{\beta} = \tilde{\beta}^c T''_c \quad (c = 1,
\ldots, \dim {\cal G} - 1)$. With this
change Virasoro transformations appear both
before and after the gauge-fixing.
However, the transformation laws of the remnant fields after the
gauge-fixing are generally different from the original ones.

In order to examine the Virasoro transformations of the gauge field
$\Lambda$, we decompose it as
$$
\Lambda=\sum_\gamma \Lambda^\gamma E_\gamma+\sum_\alpha
\Lambda^\alpha H_\alpha,
$$
where $\{E_\gamma,H_\alpha\}$ form a Cartan-Weyl basis of the Lie
algebra ${\cal G}$.
The zero-curvature condition together with the definition \bref{change
para} produces the following Virasoro transformations before the
gauge-fixing:
\bea
\nonumber
\delta \Lambda^\gamma&=&\eps
\dot{\Lambda}^\gamma+(1+\sum_\alpha
(\alpha,\gamma)\tilde k_\alpha)\dot{\eps}\Lambda^\gamma,
\\
\label{lag dif tra}
\delta \Lambda^\alpha&=&\eps\dot{\Lambda}^\alpha+\dot{\eps}
\Lambda^\alpha +\tilde k_\alpha\ddot{\eps}.
\eea
Notice that the fields living on the Cartan subalgebra (indices
$\alpha$) transform
as weight-one tensors, generally with an inhomogeneous extension term.
Instead the fields living in the root spaces (indices $\gamma$)
transform as tensors of weight $ 1+\sum_\alpha
(\alpha,\gamma)\tilde k_\alpha$.
Had we considered a change of parameters \bref{change para} without the
$\dot{\epsilon} H$ term then all fields would have transformed as
weight-one
tensors. This is equivalent to taking the usual Sugawara energy-momentum
tensor as the generator of the Virasoro transformations in the Kac-Moody
Hamiltonian reduction framework. Then the addition of the
$\dot{\epsilon} H$ term in \bref{change para} corresponds to changing
the
realization of the Virasoro group by considering an improved Sugawara
energy-momentum tensor as the generator.

The transformations generated by $\eps$ remain to be
Virasoro transformations after the gauge-fixing
procedure. Indeed, the parameter $\eps$ lives in the subspace
generated by $E_+$ so it is one of remnant parameter (we can take the
$T''_c$ Lie algebra elements as the generators of all the ${\cal
G}^{(i)}_{j,m}$ subspaces except the one of
${\cal G}_{\underline{1},1}$, i.e. $E_+$).
The zero-curvature condition \bref{zeroc} after the gauge-fixing
reads:
\bea
\label{zcexp}
[M,\tilde{\beta}]&=&\dot{\tilde{\beta}}-\delta W
-[W,\tilde{\beta}]+\eps \dot{W} + \dot{\eps}(M+W)
+ \ddot{\eps} H+\dot{\eps}[H,M+W].
\eea
To solve the zero-curvature condition we have to expand $H$ and $W$
as follows:
\bea
\label{h exp k}
H&=&k_0 h+\sum_ik_iH_i+\sum_\sigma k_\sigma H_\sigma,
\\\nonumber
W&=&\sum_iW^iH_i+\sum_\alpha W^\alpha e_\alpha +\sum_\rho W^\rho e_\rho.
\eea
In the expansion of $H$, $h$ is the sl(2)-embedding element, $\{H_i\}$
span ${\cal G}_W\cap{\cal H}$ and $\{H_\sigma\}$ form a basis of the
rest of ${\cal H}$.
In the expansion of $W$, $W^i$ are the fields living in $\cal H$,
$W^\alpha$ are the fields living in $\hat{\cal G}_0$ but not in $\cal H$
and $W^\rho$ are the rest of remnant fields.
The following relations hold:
\bea
\nonumber
&[h,e_\rho]=-j_{(\rho)}e_\rho,\quad\quad\quad[h,e_\alpha]=0,
\\\nonumber
&[H_i,e_\rho]=r_{i(\rho)}e_\rho,\quad\quad
\quad[H_i,e_\alpha]=r_{i(\alpha)}e_\alpha.
\eea

One can study the propagation of the parameter $\epsilon$ through
the equations imposed by the zero-curvature condition at each
level in the gradation of $\cal G$.
The result of this analysis is:
\be
\label{propeps}
\tilde{\beta} = -(1+k_0)\dot{\epsilon} h - \dot{\epsilon} k_{\sigma}
H_{\sigma} - \ddot{\epsilon} E_- + {\rm \ (terms \ without \ }
\epsilon).
\ee
Once we introduce \bref{propeps} in \bref{zcexp} we get the residual
infinitesimal transformations of the remnant fields, $\delta W$, under
the $\epsilon$ sector.
There are some cancellations due to the presence of
the term $\ddot{\eps} H$ which cut off the propagation of the $k_0$ and
$k_\sigma$ parameters.
Hence the only surviving arbitrariness comes from the $k_i$ parameters.
The result is summarized as:
\begin{itemize}
\item
The field $T$ living in the subspace generated by $E_-$,
which is one of the $e_\rho$ generators,
transforms as a quasi-primary field of weight two:
\be
\delta T = \epsilon \dot{T} + 2 \dot{\epsilon} T -
\dddot{\epsilon}.
\ee

\item
Fields living in the subspace spanned by $H_i$, $W^i$, transform as
weight-one fields plus a term $\ddot{\epsilon}$:
\be
\delta W^i = \epsilon \dot{W^i} + \dot{\eps} W^i + k_i
\ddot{\eps},\quad \quad (i=1,\ldots,\dim {\cal H} \cap {\cal
G}_W).
\ee

\item
The rest of remnant fields living in $\hat{\cal G}_{m=-j}$,
$W^\rho$ and $W^\alpha$, are primary fields:
\bea
\nonumber
\delta W^\rho = \eps \dot{W}^\rho+ \left( 1 +
j_{(\rho)} + \sum_i k_i r_{i(\rho)} \right)\dot{\eps} W^\rho,
\\
\delta W^\alpha = \eps \dot{W}^\alpha + \left( 1 +
\sum_i k_i r_{i(\alpha)} \right)\dot{\eps} W^\alpha.
\eea

\end{itemize}

In general, the field living in $\hat{\cal G}_{m=-j}$ has weight
$1+j$ apart from possible shifts, which exist in case
the subspace ${\cal H} \cap {\cal G}_W$ is non-trivial.
The following relation holds:
\be
\sum {\rm weights} := \sum_{j \geq 0} n_j (1 + j) = \frac12 \left( \dim
{\cal G} + N({\cal G}) \right).
\ee
There is no explicit general formula for the transformations generated
by the other remnant parameters. They are precisely specific chiral
${\cal W}$-transformations because we have a set of infinitesimal
transformations with closed algebra and containing a Virasoro sector
with the weight-two quasi-primary field $T$.

\subsection{Inequivalent sl$(2,R)$ embeddings}
\hspace{\parindent}%

It is useful to separate the set of all possible sl$(2,R)$ embeddings in
$\cal{G}$ into classes of equivalent embeddings. Two embeddings ${\cal
S}_1$
and ${\cal S}_2$ are said to be equivalents if there exists an
automorphism of $\cal{G}$ mapping ${\cal S}_1$ onto ${\cal S}_2$.
There will be as many admissible gauge-fixings \bref{pgf} as
classes of equivalent sl$(2,R)$ embeddings.

Given a canonical decomposition of $\cal{G}$ (i.e. given a Cartan
subalgebra
of $\cal{G}$, $\cal{H}$, a set of positive roots, $\Delta_+$, and a set of
simple roots, $\Pi$),
\be
\nonumber
{\cal G} = \sum_{\alpha \in \Delta_+} \dot{+} \ {\cal G}_{-\alpha}  \
\dot{+} \ {\cal H} \ \dot{+}
\sum_{\alpha \in \Delta_+} \dot{+} \ {\cal G}_{\alpha},
\ee
and a sl$(2,R)$ embedding in ${\cal G}$, ${\cal S}$,
we can always choose a member of the same class of equivalence of
$\cal{S}$ such that:
\be
\nonumber
h \in {\cal H}, \quad \quad  {\rm i.e.} \quad h = H_{\delta} \quad
{\rm where} \quad
\delta = \sum_{\beta \in \Pi} c_{\beta} \beta,
\ee
\be
\label{E+-}
E_{\pm} = \sum_{\gamma \in {\Gamma}_{\delta}} e_{\pm \gamma}, \quad
\quad \quad e_{\gamma} \in {\cal G}_{\gamma}, \quad \quad \quad
{\Gamma}_{\delta} = \left\{ \gamma \in \Delta_+ \mid \left(\gamma,
\delta \right) = 1 \right\};
\ee
$\delta$ is the {\it defining vector} of such an embedding. Let us
consider the
Dynkin diagram of $\cal{G}$. We construct the {\it characteristic} of
this
$sl(2,R)$ embedding by writing down the number $\left( \beta, \delta
\right)$
under the dot of the Dynkin which represents the root $\beta$ for each
$\beta \in \Pi$. Two important results follow \cite{Dyn,Rag}:
\begin{itemize}
\item Two sl$(2,R)$ embeddings are equivalent if and only if their
      characteristics coincide.
\item If a characteristic is associated with a sl$(2,R)$ embedding then
      it exhibits numbers of the set $\left\{ 0, \frac12, 1 \right\}$.
\end{itemize}
It can be shown that the potential characteristic which exhibits a $1$
under
every dot always gives rise to a sl$(2,R)$ embedding, which is known
as  the {\it principal} sl$(2,R)$ embedding.

\medskip

As an example we present here the case ${\cal G} = {\rm sp}(4,R)$.
The Dynkin diagram, normalizations and positive roots set for
sp$(4,R)$ are:

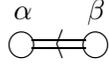
\begin{picture}(200,40)(-50,0)
\put(105.5,7.5){$\bigcirc$}
\put(133.5,7.5){$\bigcirc$}
\put(115,9){\line(1,0){20}}
\put(115,12){\line(1,0){20}}
\put(123,8){$\langle$}
\put(108,20){$\alpha$}
\put(136,20){$\beta$}
\end{picture}
$$
(\alpha, \alpha) = \frac{1}{6} \quad \quad
(\beta, \beta) = \frac{1}{3} \quad \quad
(\alpha, \beta) = - \frac{1}{6}, \quad \quad \quad
\Delta_+ = \left\{ \alpha, \beta, \alpha + \beta, 2\alpha + \beta \right\}.
$$

There are only three classes of non-equivalent sl$(2,R)$ embeddings.
Their characteristics are:

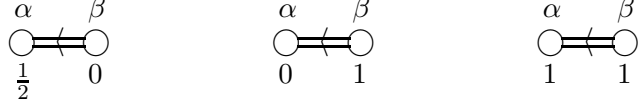
\begin{picture}(300,40)(-50,0)
\put(5.5,17.5){$\bigcirc$}
\put(33.5,17.5){$\bigcirc$}
\put(15,19){\line(1,0){20}}
\put(15,22){\line(1,0){20}}
\put(23,18){$\langle$}
\put(8,30){$\alpha$}
\put(36,30){$\beta$}
\put(8,3){$\frac12$}
\put(36,5){0}

\put(105.5,17.5){$\bigcirc$}
\put(133.5,17.5){$\bigcirc$}
\put(115,19){\line(1,0){20}}
\put(115,22){\line(1,0){20}}
\put(123,18){$\langle$}
\put(108,30){$\alpha$}
\put(136,30){$\beta$}
\put(108,5){0}
\put(136,5){1}

\put(205.5,17.5){$\bigcirc$}
\put(233.5,17.5){$\bigcirc$}
\put(215,19){\line(1,0){20}}
\put(215,22){\line(1,0){20}}
\put(223,18){$\langle$}
\put(208,30){$\alpha$}
\put(236,30){$\beta$}
\put(208,5){1}
\put(236,5){1}
\end{picture}

Our matrix conventions for the generators of sp$(4,R)$ are:
\bea
\nonumber
& H_{\alpha}=\frac{1}{12}\left(\begin{array}{cccc}
                              1 & 0 & 0 & 0 \\
                              0 & -1 & 0 & 0 \\
                              0 & 0 & -1 & 0 \\
                              0 & 0 & 0 & 1 \end{array}\right)\quad\quad
H_{\beta}=\frac{1}{6}\left(\begin{array}{cccc}
                              0 & 0 & 0 & 0 \\
                              0 & 1 & 0 & 0 \\
                              0 & 0 & 0 & 0 \\
                              0 & 0 & 0 & -1 \end{array}\right)
\\\nonumber
& E_{\alpha}=\frac{1}{\sqrt{12}}\left(\begin{array}{cccc}
                              0 & 1 & 0 & 0 \\
                              0 & 0 & 0 & 0 \\
                              0 & 0 & 0 & 0 \\
                              0 & 0 & -1 & 0
\end{array}\right)\quad\quad
E_{-\alpha}=\frac{1}{\sqrt{12}}\left(\begin{array}{cccc}
                              0 & 0 & 0 & 0 \\
                              1 & 0 & 0 & 0 \\
                              0 & 0 & 0 & -1 \\
                              0 & 0 & 0 & 0 \end{array}\right)
\\\nonumber
& E_{\beta}=\frac{1}{\sqrt{6}}\left(\begin{array}{cccc}
                              0 & 0 & 0 & 0 \\
                              0 & 0 & 0 & 1 \\
                              0 & 0 & 0 & 0 \\
                              0 & 0 & 0 & 0 \end{array}\right)\quad\quad
E_{-\beta}=\frac{1}{\sqrt{6}}\left(\begin{array}{cccc}
                              0 & 0 & 0 & 0 \\
                              0 & 0 & 0 & 0 \\
                              0 & 0 & 0 & 0 \\
                              0 & 1 & 0 & 0 \end{array}\right)
\\\nonumber
& E_{\alpha+\beta}=\frac{1}{\sqrt{12}}\left(\begin{array}{cccc}
                              0 & 0 & 0 & 1 \\
                              0 & 0 & 1 & 0 \\
                              0 & 0 & 0 & 0 \\
                              0 & 0 & 0 & 0 \end{array}\right)\quad\quad
E_{-(\alpha+\beta)}=\frac{1}{\sqrt{12}}\left(\begin{array}{cccc}
                              0 & 0 & 0 & 0 \\
                              0 & 0 & 0 & 0 \\
                              0 & 1 & 0 & 0 \\
                              1 & 0 & 0 & 0 \end{array}\right)
\\\nonumber
& E_{2\alpha+\beta}=\frac{1}{\sqrt{6}}\left(\begin{array}{cccc}
                              0 & 0 & 1 & 0 \\
                              0 & 0 & 0 & 0 \\
                              0 & 0 & 0 & 0 \\
                              0 & 0 & 0 & 0 \end{array}\right)\quad\quad
E_{-(2\alpha+\beta)}=\frac{1}{\sqrt{6}}\left(\begin{array}{cccc}
                              0 & 0 & 0 & 0 \\
                              0 & 0 & 0 & 0 \\
                              1 & 0 & 0 & 0 \\
                              0 & 0 & 0 & 0 \end{array}\right).
\eea

Here we display a representative of every class of equivalent embeddings and
the corresponding branchings:

\begin{itemize}

\item \underline{$(0,1)$ embedding}:
\bea
\nonumber
& h = 6 H_{\alpha} + 6 H_{\beta} \quad \quad \quad
E_{\pm} = \sqrt{6} E_{\pm (\alpha + \beta)}
\\\nonumber
& {\cal G} = {\cal G}^{(1)}_{\underline 1} \ \dot{+} \ {\cal G}^{(2)}_1 \
              \dot{+} \ {\cal G}^{(3)}_1 \ \dot{+} \ {\cal G}_0
\\\nonumber
& {\cal G}^{(2)}_1 = \langle E_{2\alpha + \beta}, E_{\alpha}, E_{-\beta}
\rangle \quad\quad
{\cal G}^{(3)}_1 = \langle E_{\beta}, E_{-\alpha}, E_{-(2\alpha + \beta)}
\rangle
\\
\label{emb(0,1)2}
& {\cal G}_0 = \langle H_{\alpha} \rangle.
\eea

\item \underline {$(\frac12,0)$ embedding}:
\bea
\nonumber
& h = 6 H_{\alpha} + 3 H_{\beta} \quad \quad \quad
E_{\pm} = \sqrt{3} E_{\pm (2\alpha + \beta)}
\\\nonumber
& {\cal G} = {\cal G}_{\underline 1} \ \dot{+} \ {\cal G}^{(1)}_{\frac12} \
           \dot{+} \ {\cal G}^{(2)}_{\frac12} \ \dot{+} \ {\cal G}^{(1)}_0 \
           \dot{+} \ {\cal G}^{(2)}_0 \ \dot{+} \ {\cal G}^{(3)}_0
\\\nonumber
& {\cal G}^{(1)}_{\frac12} = \langle E_{\alpha + \beta}, E_{-\alpha} \rangle
\quad \quad
{\cal G}^{(2)}_{\frac12} = \langle E_{\alpha}, E_{-(\alpha + \beta)} \rangle
\\
\label{emb(1/2,0)}
& {\cal G}^{(1)}_0 = \langle E_{\beta} \rangle
\quad \quad
{\cal G}^{(2)}_0 = \langle E_{-\beta} \rangle
\quad \quad
{\cal G}^{(3)}_0 = \langle H_{-\beta} \rangle.
\eea

\item \underline{$(1,1)$ (principal) embedding}:
\bea
\nonumber
& h = 18 H_{\alpha} + 12 H_{\beta} \quad \quad \quad
E_{\pm} = \sqrt{18} E_{\pm \alpha} + \sqrt{12} E_{\pm \beta}
\\\nonumber
& {\cal G} = {\cal G}_{\underline 1} \ \dot{+} \ {\cal G}_3
\\\nonumber
& {\cal G}_3 = \langle E_{2\alpha + \beta}, E_{\alpha + \beta},
                     \sqrt{3} E_{\beta} - \sqrt{2} E_{\alpha},
                     H_{\alpha} - H_{\beta},
                      \sqrt{3} E_{-\beta} - \sqrt{2} E_{-\alpha},
\\
\label{emb(1,1)no}
& \,\,\,\,\,\,\,\,\,\,\,\,\,\,\,
E_{-(\alpha + \beta)}, E_{-(2\alpha + \beta)} \rangle.
\eea
We consider another element of this conjugacy class because
the previous one
produces a gauge fixing such that there is no gauge-fixed Lagrangian
in terms of coordinates and velocities:
\bea
\nonumber
& h = 18 H_{\alpha} + 6 H_{\beta} \quad \quad \quad
E_{\pm} = \sqrt{18} E_{\pm (\alpha+\beta)} + \sqrt{12} E_{\mp \beta}
\\\nonumber
& {\cal G} = {\cal G}_{\underline 1} \ \dot{+} \ {\cal G}_3
\\\nonumber
& {\cal G}_3 = \langle E_{2\alpha + \beta}, E_{\alpha},
                     \sqrt{3} E_{-\beta} - \sqrt{2} E_{\alpha + \beta},
                     H_{\alpha} + 2 H_{\beta},
                      \sqrt{3} E_{\beta} - \sqrt{2} E_{-(\alpha+\beta)},
\\
\label{emb(1,1)}
& \,\,\,\,\,\,\,\,\,\,\,\,\,\,\, E_{-\alpha}, E_{-(2\alpha + \beta)} \rangle.
\eea
\end{itemize}

\bigskip

For completeness we also present here a representative of each of the
two classes of equivalent sl$(2,R)$ embeddings in sl$(3,R)$:

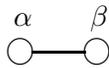
\begin{picture}(200,40)(-50,0)
\put(105,7.5){$\bigcirc$}
\put(134,7.5){$\bigcirc$}
\put(115,10){\line(1,0){20}}
\put(108,20){$\alpha$}
\put(137,20){$\beta$}
\end{picture}
$$
(\alpha, \alpha) = (\beta, \beta) = \frac{1}{3} \quad \quad
(\alpha, \beta) = - \frac{1}{6}, \quad \quad \quad
\Delta_+ = \left\{ \alpha, \beta, \alpha + \beta \right\}.
$$

\begin{itemize}
\item \underline{principal $(1,1)$ embedding}:
\bea
\nonumber
& h = 6 H_{\alpha} + 6 H_{\beta} \quad \quad \quad
E_{\pm} = \sqrt{6} E_{\pm \alpha} + \sqrt{6} E_{\pm \beta}
\\\nonumber
& {\cal G} = {\cal G}_{\underline 1} \ \dot{+} \ {\cal G}_{2}
\\
& {\cal G}_{2} = \langle E_{\alpha + \beta}, E_{\beta} - E_{\alpha},
H_{\alpha} - H_{\beta}, E_{-\alpha} - E_{-\beta}, E_{-(\alpha + \beta)}
\rangle
\eea

\item \underline{non-principal $(\frac12,\frac12)$ embedding}:
\bea
\nonumber
& h = 3 H_{\alpha} + 3 H_{\beta} \quad \quad \quad
E_{\pm} = \sqrt{3} E_{\pm (\alpha + \beta)}
\\\nonumber
& {\cal G} = {\cal G}_{\underline 1} \ \dot{+} \ {\cal G}^{(1)}_{\frac12} \
           \dot{+} \ {\cal G}^{(2)}_{\frac12} \ \dot{+} \ {\cal G}_0
\\\nonumber
& {\cal G}^{(1)}_{\frac12} = \langle E_{\alpha}, E_{-\beta} \rangle
\quad \quad
{\cal G}^{(2)}_{\frac12} = \langle E_{\beta}, E_{-\alpha} \rangle
\quad \quad
& {\cal G}_0 = \langle H_{\alpha} - H_{\beta} \rangle.
\eea
\end{itemize}

The sl$(3,R)$ subalgebra of sp$(6,R)$ that we have considered in
sect.\,5 is realized by taking the following subset of the
$\phi_{A_{ij}}$ quadratic constraints in the $M=3$ case:
\bea
\nonumber
& E_{\alpha} = \frac{1}{\sqrt{6}} p_1 p_2, \quad\quad
E_{\beta} = \frac{1}{\sqrt{6}} p_3 x_2, \quad\quad
E_{\alpha+\beta} = -\frac{1}{\sqrt{6}} p_1 p_3,
\\\nonumber
& E_{-\alpha} = -\frac{1}{\sqrt{6}} x_1 x_2, \quad\quad
E_{-\beta} = \frac{1}{\sqrt{6}} p_2 x_3, \quad\quad
E_{-(\alpha+\beta)} = \frac{1}{\sqrt{6}} x_1 x_3,
\\\nonumber
& H_{\alpha} = \frac16 (p_2 x_2 + p_1 x_1),\quad\quad\quad
H_{\beta} = \frac16 (p_3 x_3 - p_2 x_2).
\eea

\bigskip


\section{Diffeomorphism invariance and finite transformations of the
Sp$(2M)$ model}
\hspace{\parindent}%

Here we show first the invariance of the Sp$(2M)$ model under ordinary
diffeomorphism transformations, along the lines of Appendix A.
We shall later present the model's gauge symmetry transformations in
their finite form.

According to Appendix A we can perform the change of parameters
\bref{change para} with $H = \sum_{i=1}^M \tilde{k}_{\alpha_i} H_{\alpha_i},$
where $\alpha_i$ are the simple roots of sp$(2M,R)$ and
$\tilde{k}_{\alpha_i}$ are constants.
When ${\cal G}$=sp$(2M,R)$ then $H$ is the diagonalized matrix:
$$
H= \left( \begin{array}{cc} N & 0 \\
                            0 & -N^{\top} \end{array} \right),
$$
\be
\label{matrixN}
N= \frac{1}{4(M+1)} \left( \begin{array}{ccccc}
          \tilde{k}_{\alpha_1} & & & & 0 \\
          & \tilde{k}_{\alpha_2} - \tilde{k}_{\alpha_1} & & & \\
          & & \ddots & & \\
          & & & \tilde{k}_{\alpha_{M-1}}-\tilde{k}_{\alpha_{M-2}} & \\
          0 & & & & 2 \tilde{k}_{\alpha_M}-\tilde{k}_{\alpha_{M-1}}
\end{array} \right).
\ee
where $\alpha_M$ is the longest root.

As stated in appendix A (see \bref{lag dif tra}), the Lagrange
multipliers
transform as primary fields with, eventually, $\ddot{\eps}$ terms under
the $\eps$ sector infinitesimal transformations. For the matter and
auxiliary variables, this change of parameters produces the following
infinitesimal transformations:
$$
\tilde{\delta}_{\eps}r = \eps \dot{r} + \dot{\eps} N r + \eps A F,
$$
$$
\tilde{\delta}_{\eps}F = -\eps \dot{F} - \dot{\eps} (F + NF) +
\eps \left( A^{-1}BAF - A^{-1}\dot{A}F-B^{\top}F-\dot{K} - B^{\top}K -
Cr \right).
$$
These transformations are equivalent to diffeomorphism transformations.
In order to show this let us introduce
an antisymmetric combination of the equations of motion:
\bea
\nonumber
& \delta_{\eps} q^i (t) = \tilde{\delta} q^i (t) + \int \dif t'
M^{ij}(t,t') [L]_{q^j} (t'),
\\\nonumber
&{\rm where} \quad\quad \left\{ \begin{array}{ll} q^i = x_i &
                                      (i=1,\ldots,M) \\
                                      q^i = F_i & (i=M+1,\ldots,2M)
\end{array} \right.
\\\nonumber
&{\rm and} \quad\quad\quad M(t,t')=\left( \begin{array}{cc} 0 &
                           \eps(t)\delta(t-t')I \\
                           -\eps(t)\delta(t-t')I & \bar{M}(t,t')
\end{array} \right),
\\\nonumber
&\bar{M}(t,t') = -\eps(t) \left( B^{\top}(t)
A^{-1}(t) - A^{-1}(t) B(t) \right) \delta(t-t') +
\\
\label{ineqmot diff}
&\eps(t') A^{-1}(t')\frac{\dif}{\dif t'} \delta(t-t')-
\eps(t) A^{-1}(t)\frac{\dif}{\dif t} \delta(t-t').
\eea
It can be shown that
$$
M^{\top}(t',t)=-M(t,t');
$$
so $\delta_{\eps}$ is a symmetry transformation of the action too and
\be
 \delta_{\eps}r=\eps \dot{r} + \dot{\eps} Nr,
\hskip 7mm
 \delta_{\eps}F=\eps \dot{F} - \dot{\eps} NF,
\ee
which are diffeomorphism transformations for the matter and auxiliary
variables. They transform as primary fields.

In summary, the infinitesimal gauge transformations of the
Sp$(2M)$ model before the gauge-fixing are:

$\bullet$ Diffeomorphism transformations:

\bea
\nonumber
&\delta \Lambda^\gamma=\eps
\dot{\Lambda}^\gamma+(1+\sum_\alpha
(\alpha,\gamma)\tilde k_\alpha)\dot{\eps}\Lambda^\gamma, \quad\quad\quad
\delta \Lambda^\alpha=\eps\dot{\Lambda}^\alpha+\dot{\eps}
\Lambda^\alpha +\tilde k_\alpha\ddot{\eps},
\\\nonumber
& \delta_{\eps}r=\eps \dot{r} + \dot{\eps} Nr,
\quad\quad\quad
\delta_{\eps}F=\eps \dot{F} - \dot{\eps} NF,
\eea

$\bullet$ Yang-Mills type transformations:

\bea
\nonumber
&\delta\Lambda=\dot{\tilde{\beta}}-[\Lambda,\tilde{\beta}],
\quad\quad\quad \delta r={\tilde{\beta}}_A (K+F) + {\tilde{\beta}}_B r,
\\\nonumber
&\delta F=-A^{-1}\left[{\tilde{\beta}}_A(\dot{K}+\dot{F})+
{\tilde{\beta}}_A B^\top K +
{\tilde{\beta}}_A Cr - B {\tilde{\beta}}_A F +
{\dot{\tilde{\beta}}}_A F \right] - {\tilde{\beta}}_B^\top F,
\eea

where:

$$
\Lambda=
\sum_\gamma \Lambda^\gamma E_\gamma+\sum_\alpha
\Lambda^\alpha H_\alpha=
\left(\begin{array}{cc}
              B&A\\
             -C&-B^\top
              \end{array}\right), \quad\quad\quad
\tilde{\beta}=\left(\begin{array}{cc}
              {\tilde{\beta}}_B&{\tilde{\beta}}_A\\
             -{\tilde{\beta}}_C&-{\tilde{\beta}}^{\top}_B
              \end{array}\right).
$$

\vskip 2mm

After performing the gauge-fixing, Lagrange multipliers still transform
as primary or quasi-primary fields (see appendix A) whereas matter and
auxiliary fields do not have, in general, a nice behavior under Diff
transformations. For instance,
if the $E_-$ element of the sl$(2,R)$ embedding is
taken to live in the $A$ or $B$ sectors of a general sp$(2M,R)$ matrix,
then non-desired $\ddot{\epsilon}$ terms appear in the residual
$\epsilon$ transformations coming from the algorithm described in
appendix A (see \bref{propeps}) through the ${\tilde{\beta}}_A$ or
${\tilde{\beta}}_B$ factors of \bref{mat tra} and \bref{aux tra}.
In any case, the only undetermined constants that remain in the
infinitesimal transformations after the gauge-fixing from those in the
\bref{h exp k} decomposition of $H$ are the constants $k_i$
as in the residual transformations for the Lagrange multipliers.

\vskip 5mm

Finite trans\-form\-ations%
\footnote{
For a recent discussion on finite gauge transformations see \cite{GPR}.}
can be obtained by exponentiating the
infinitesimal ones as
$
{X^i}'=\exp\{\theta^\alpha\Gamma_\alpha\}X^i,
$
where the generators $\Gamma_\alpha=R^i_\alpha\frac{\partial}
{\partial X^i}$ satisfy
$
[\Gamma_\alpha,\Gamma_\beta]=f^\gamma_{\alpha\beta}\Gamma_\gamma
$
and $ X^i$ represents any of the variables.
The coefficients $f^\gamma_{\alpha\beta}$ are the structure functions of
the sp$(2M)$ gauge algebra.

It is useful to perform the integration using the matrix notation.
The explicit form of the finite gauge transformations is considered
in the following four sets of transformations. Any finite transformations
may be obtained by the composition of them.
\vskip 3mm

$\bullet$ The diffeomorphism transformations:
\bea
\nonumber
&\Lambda'^{\gamma} (t) = \dot{f}(t)^{1+\sum_{\alpha} (\alpha,\gamma)
\tilde{k}_{\alpha}} \Lambda^{\gamma} (f(t)), \quad\quad\quad
\Lambda'^{\alpha} (t) = \dot{f}(t) \Lambda^{\alpha} (f(t)) +
\tilde{k}_{\alpha} \frac{\ddot{f}(t)}{\dot{f}(t)},
\\
&r'_i (t) = \dot{f}(t)^{N_{ii}} r(f(t)), \quad\quad\quad
F'_i (t) = \dot{f}(t)^{-N_{ii}} F(f(t)), \quad\quad\quad
i=1,\ldots,M.
\eea
where $N=(N_{ij})$ is the diagonalized constant matrix given in eq.
\bref{matrixN}.

$\bullet$ Transformations generated by ${\tilde{\beta}}_A$:
$$
A'=A+\{{\dot{\tilde{\beta}}}_A-{\tilde{\beta}}_A
B^\top-B{\tilde{\beta}}_A\}+{\tilde{\beta}}_A C {\tilde{\beta}}_A,
\hskip 7mm
 B'=B-{\tilde{\beta}}_A C,\quad\quad\quad C'=C,
$$
\be
r'=r+{\tilde{\beta}}_A (K+F),
\hskip 7mm
F'={A'}^{-1}\left[AF-{\tilde{\beta}}_A\{\partial_t(K+F)+
B^\top(K+F)+Cr\}\right].
\ee

\vskip 3mm

$\bullet$ Transformations generated by ${\tilde{\beta}}_B$:
$$
A'=e^{{\tilde{\beta}}_B}Ae^{{\tilde{\beta}}^{\top}_B},\,\,\,\,
B'=e^{{\tilde{\beta}}_B}(B-\partial_t)e^{-{\tilde{\beta}}_B},\,\,\,\,
C'=e^{-{\tilde{\beta}}^{\top}_B}Ce^{-{\tilde{\beta}}_B},
$$
\be
r'=e^{{\tilde{\beta}}_B}r,\,\,\,\,
F'=e^{-{\tilde{\beta}}^{\top}_B}F.
\ee
\vskip 3mm

$\bullet$ Transformations generated by ${\tilde{\beta}}_C$:
$$ A'=A,\quad\quad\quad r'=r,\quad\quad\quad F'=F,
$$
\be B'=B+A {\tilde{\beta}}_C,
\hskip 7mm
C'=C+\{{\dot{\tilde{\beta}}}_C+ {\tilde{\beta}}_C B+B^\top
{\tilde{\beta}}_C\}+{\tilde{\beta}}_C A {\tilde{\beta}}_C.
\ee
\vskip 3mm


\end{document}